\documentclass[%
 reprint,
 superscriptaddress, 
 amsmath,amssymb,
 aps,
 pra,
 floatfix,
]{revtex4-2}

\usepackage{graphicx}
\usepackage{dcolumn}
\usepackage{bm}
\usepackage{dsfont}
\usepackage{xcolor}
\usepackage{enumitem}
\usepackage{multirow}
\usepackage{hyperref}
\usepackage{array}
\usepackage{float}
\usepackage{csquotes}
\usepackage[T1]{fontenc}

\MakeOuterQuote{"}

\usepackage[textsize=tiny]{todonotes}

\usepackage{listings}

\definecolor{customcolorblue}{HTML}{4573ae}

\begin{document}

\lstdefinelanguage{Julia}{
  morekeywords={
    abstract,break,case,catch,const,continue,do,else,elseif,end,begin,
    export,false,for,function,global,if,import,let,local,macro,module,
    mutable,primitive,quote,return,struct,true,try,type,using,while,
    in,isa,where,nothing
  },
  sensitive=true,
  morecomment=[l]\#,
  morestring=[b]",
  morestring=[b]',
  morekeywords=[2]{@async,@time,@resumable,@inline,@inbounds,@views,@process,@yield},
}

\lstdefinestyle{julia}{
  language=Julia,
  basicstyle=\ttfamily\scriptsize,
  keywordstyle=\color{blue}\bfseries,
  keywordstyle=[2]\color{magenta}, 
  commentstyle=\color{gray}\itshape,
  stringstyle=\color{orange},
  showstringspaces=false,
  breaklines=true,
  columns=flexible,
  keepspaces=true,
  frame=single,
  rulecolor=\color{black!30},
  numbers=left,
  numberstyle=\tiny\color{gray},
  xleftmargin=0.5em,
  framexleftmargin=0em,
  numbersep=4pt,
}

\lstset{style=julia}

\title{QuantumSavory: Write Symbolically, Run on Any Backend -- A Unified Simulation Toolkit for Quantum Computing and Networking}
\title{QuantumSavory: symbolic modeling and multi-backend simulation of quantum computing and networking systems}

\author{Hana Kimlee}
\thanks{These authors contributed equally to this work.}
\affiliation{NSF-ERC Center for Quantum Networks, The University of Arizona, Tucson, AZ 85721, USA}

\author{Leonardo Bacciottini}
\thanks{These authors contributed equally to this work.}
\affiliation{(Corresponding author: lbacciottini@umass.edu)}
\affiliation{NSF-ERC Center for Quantum Networks, The University of Arizona, Tucson, AZ 85721, USA}
\affiliation{College of Information and Computer Sciences, University of Massachusetts Amherst, USA}

\author{Abhishek Bhatt}
\affiliation{NSF-ERC Center for Quantum Networks, The University of Arizona, Tucson, AZ 85721, USA}
\affiliation{College of Information and Computer Sciences, University of Massachusetts Amherst, USA}

\author{Andrew Kille}
\affiliation{Center for Computational Quantum Physics, Flatiron Institute, New York, NY, USA}
\affiliation{Department of Physics, New York University, New York, NY, USA}

\author{Stefan Krastanov}
\email{skrastanov@umass.edu}
\affiliation{NSF-ERC Center for Quantum Networks, The University of Arizona, Tucson, AZ 85721, USA}
\affiliation{College of Information and Computer Sciences, University of Massachusetts Amherst}
\affiliation{Department of Physics, University of Massachusetts Amherst, USA}


















\begin{abstract}

  Progress in quantum computing and networking depends on \emph{codesign} across abstraction layers: device-level noise and heterogeneous hardware, algorithmic structure, and distributed classical control. We present QuantumSavory, an open-source toolkit built to make such end-to-end studies practical by cleanly separating a \emph{symbolic computer-algebra frontend} from interchangeable numerical simulation backends. States, operations, measurements, and protocol logic are expressed in a backend-agnostic symbolic language; the same model can be executed across multiple backends (e.g., stabilizer, wavefunction, phase-space), enabling rapid exploration of accuracy-performance tradeoffs without rewriting the model. Furthermore, new custom backends can be added via a small, well-defined interface that immediately reuses existing models and protocols.

  QuantumSavory also addresses the classical-quantum interaction inherent to LOCC protocols via discrete-event execution and a tag/query system for coordination. Tags attach structured classical metadata to quantum registers and message buffers, and queries retrieve, filter, or wait on matching metadata by wildcards or arbitrary predicates. This yields a data-driven control plane where protocol components coordinate by publishing and consuming semantic facts (e.g., resource availability, pairing relationships, protocol outcomes) rather than by maintaining rigid object graphs or bespoke message plumbing, improving composability and reuse as models grow. Our toolkit is also not limited to qubits and Bell pairs; rather, any networking dynamics of any quantum system under any type of multipartite entanglement can be tackled. Lastly, QuantumSavory ships reusable libraries of standard states, circuits, and protocol building blocks with consistent interfaces, enabling full-stack examples to be assembled, modified, and compared with minimal glue code. 
\end{abstract}

\begin{abstract}
We present QuantumSavory, an open-source toolkit for quantum systems simulation,  designed to separate a symbolic computer-algebra frontend from interchangeable numerical backends. States, operations, measurements, and protocol logic are expressed in a backend-agnostic symbolic language; the same model can be executed across multiple backends (e.g., stabilizer, wavefunction, phase-space), enabling seamless exploration of accuracy-performance tradeoffs. New backends can be added via a small, well-defined interface.

QuantumSavory also addresses the classical-quantum interaction inherent to LOCC via discrete-event simulation and a tag/query mechanism for coordination: protocols interact by publishing and consuming semantic facts (e.g., resource availability, entanglement location, outcomes), improving composability and reuse as models grow. Our toolkit is also not limited to qubits and Bell pairs; rather, any dynamics of any composite quantum system can be tackled. Lastly, QuantumSavory ships repositories of standard states, circuits, and protocols, enabling full-stack examples to be assembled, modified, and compared with minimal glue code. 

\end{abstract}

\maketitle





\section{Introduction}

Quantum Information Science (QIS) is inherently multidisciplinary, demanding expertise that ranges from the intricate experimental physics of hardware design to the theory guiding applications, and further to foundational concepts in mathematics, computability, and physics. This multidimensional complexity has resulted in a fragmented landscape where theorists and hardware developers often operate in silos. The absence of standardized tooling, machine-readable databases, and interoperable platforms stymies progress, making collaboration cumbersome and inefficient. In particular, codesign of high-level protocols and low-level hardware is incredibly difficult, made yet more challenging by the added complexity of the diversity of bespoke modeling algorithms for quantum systems (compared to classical ones).

To this end, QuantumSavory (QSavory hereafter) directly addresses these multifaceted challenges by providing an open-source ecosystem of modeling tools for quantum hardware. QSavory is implemented in Julia~\cite{bezanson2017julia}, a convenient choice given its support for high-performance numerical computing and the "multiple dispatch" language paradigm. We created QSavory to be the modeling tool that can span the entire quantum technology stack without requiring users to possess detailed knowledge of every modeling technique involved. At the same time, the framework allows users to efficiently simulate and optimize the specific subsystem or abstraction level they are working on. These abstractions and co-design capabilities enable the construction of high-fidelity digital twins, crucial for the design, validation, and optimization of quantum hardware infrastructure. 

Furthermore, different quantum processes and protocols require very different simulation strategies for efficient modeling, and QSavory unifies them by offering a symbolic, formalism-agnostic computer algebra frontend capable of expressing QIS processes, together with a simulator orchestrator that automatically selects efficient, special-purpose backends. This symbolic layer is coupled with state-of-the-art autodifferentiation capabilities, GPU acceleration, and a high-performance discrete-event simulator, enabling intricate full-stack codesign tasks.

This paper is structured as follows. Section~\ref{relatedwork} surveys related quantum network simulators, discusses the difficulties in building good digital twins, and motivates the need for QSavory. Section~\ref{background} provides background on relevant concepts in quantum information science, aimed at presenting the intricacies in deciding how to model a quantum system, the tradeoffs between accuracy and computational efficiency, and it ends with a discussion of typical quantum networking primitives. Section~\ref{qsavory} introduces the design and core abstractions of QSavory. This section presents the majority of new capabilities that QSavory offers. Section~\ref{sequence} provides a comparative evaluation against an existing simulator (SeQUeNCe), focusing on the relative ease of writing a simulation in QSavory and how the abstractions it provides make otherwise difficult tasks easy. Section~\ref{fullstack} demonstrates full-stack use cases through representative examples and serves as tutorials for the tool, giving a better sense of the edge cases one might need to worry about in a complete digital twin. Section~\ref{conclusions} concludes with a roadmap for important next steps and low hanging fruit that would be necessary for any quantum networking modeling tool to be useful for the challenging new modeling problems that are emerging.
\section{Related Work\label{relatedwork}}
\begin{figure}
        \centering
        \includegraphics[width=\linewidth]{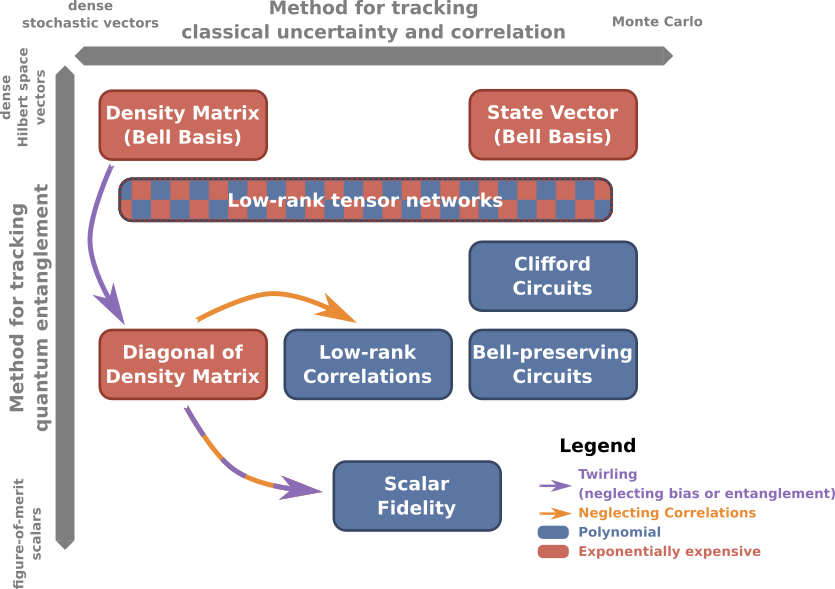}
        \caption{Two largely independent axes determine the cost--fidelity tradeoffs of classical algorithms for modeling quantum systems. The first axis concerns how \emph{classical} uncertainty and correlations are represented: one can evolve explicit, dense probability objects (e.g., exponentially large stochastic vectors or large coupled rate equations), or instead use sparse, sample-based representations such as Monte Carlo trajectories that concentrate effort on the parts of the distribution that matter; for realistic network-scale studies, the dense option is rarely the right tool (and thus the use of density matrices instead of Monte Carlo over state vectors is almost always misguided and expensive). The second axis concerns how \emph{quantum} correlations and entanglement are represented: exact state-vector or density-matrix methods incur unavoidable exponential scaling in the generic case, but specialized representations can be highly efficient for restricted dynamics (e.g.\ tensor-network methods for low-entanglement structure, or Clifford and Gaussian formalisms for restricted gate and noise sets). Pushing simplification too far can be counterproductive: extremely coarse models that track only a few summary parameters (e.g., fidelities of Werner-state pairs) are only marginally more efficient but often discard the dynamical information needed to make accurate protocol- and hardware-level decisions.}
        \label{fig:simmethods}
\end{figure}

\begin{figure*}[tb]
        \centering
        {\includegraphics[width=\linewidth]{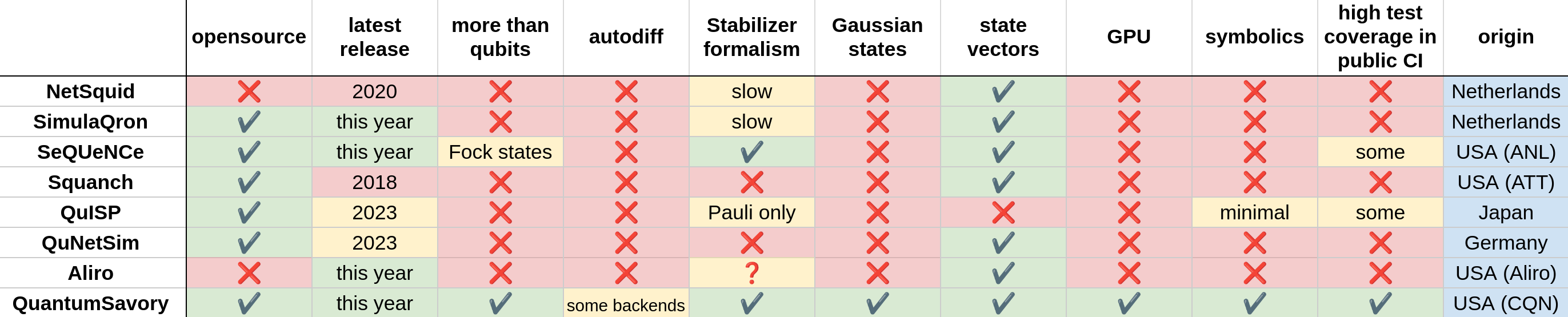}}
        \caption{Comparison to existing tools: NetSquid~\cite{coopmans2021netsquid}, SimulaQron~\cite{dahlberg2018simulaqron}, SeQUeNCe~\cite{wu2021sequence}, Squanch~\cite{bartlett2018distributed}, QuISP~\cite{satoh2022quisp}, QuNetSim~\cite{diadamo2021qunetsim}, and Aliro tools~\cite{aliro}, versus QuantumSavory. 
        All of these tools are quite capable, however we focus on the new features we consider of the greatest importance for a scalable full-stack codesign toolkit.
        }
        \label{fig:comparison}
\end{figure*}


A number of other quantum hardware simulation tools exist with a focus on networking. While they are useful tools for network science, they lack codesign and optimization capabilities and are limited to modeling a restricted set of physical systems. As can be seen from Fig.~\ref{fig:comparison}, most existing tools are only capable of modeling abstract two-level qubits without any support for more physically realistic quantum states such as multi-level systems, bosonic states, Gaussian states, or any other Hilbert space. Moreover, frequently the only supported modeling approach is the exponentially expensive state-vector formalism. QSavory’s expanded state support makes it possible to accurately model realistic quantum-networking hardware, including transducers, bosonic codes, and continuous-variable states, while seamlessly jumping between state-vector approaches and bespoke modeling techniques of only polynomial computational complexity.

Moreover, existing tools are incapable of providing gradients over simulated figures of merit, and autodifferentiation capabilities are nonexistent, making any optimization task extremely cumbersome and inefficient. In contrast, QSavory's backends offer differentiable simulation pipelines through autodifferentiation and other techniques. 

Lastly, some of these projects have seen little development in the last few years, casting doubt on whether their deficiencies will be addressed. Moreover, it is evident that none of these tools are sufficiently versatile to tackle challenging codesign problems, completely lacking any optimization capabilities, let alone autodifferentiation-based ones.

To an extent, the absence of such capabilities is the reason these tools have not seen wide deployment, even though quantum information scientists frequently express a strong desire for a full-stack simulator. Our aim with QSavory is to provide precisely that: an actively maintained tool with a symbolic frontend, a flexible backend architecture, and efficient and scalable simulation support that together enable the full-stack, optimization-capable workflow the field has been seeking.

To complement this qualitative discussion, Section~\ref{sequence} provides a direct implementation comparison between QSavory and SeQUeNCe.
\section{Background\label{background}}

This section introduces concepts from quantum information science (QIS) that are essential for understanding QSavory's design and applications. It is intended primarily for systems and networking engineers who may not have extensive backgrounds in QIS. While not self-contained, we provide brief definitions with pointers to pedagogical references for deeper exploration. Readers already familiar with quantum information science and quantum networking may skip this section.

\subsection{Quantum Systems}

Quantum information processing is often introduced through the idealized model of a closed quantum system, where states evolve unitarily under the Schrödinger equation. In this setting, information is commonly encoded in qubits, idealized two-level quantum systems that form the basic units of quantum computation and communication. 

Realistic quantum hardware, however, rarely operates in this ideal regime. Even weak coupling to uncontrolled degrees of freedom leads to open-system behavior. An open quantum system interacts with an environment, causing decoherence and dissipation. Its dynamics are non-unitary and are typically captured using quantum channels, Kraus operators, or Lindblad master equations. In this system, the Lindblad operators encode physical noise processes such as dephasing or relaxation.

Moreover, realistic quantum platforms extend beyond idealized two-level qubits. Real hardware may involve qudits, multi-level anharmonic oscillators, bosonic modes, or Gaussian states, each with distinct error mechanisms and interaction models. Accurate simulation therefore requires a representation flexible enough to capture a wide range of Hilbert spaces, noise models, and dynamical regimes.

A defining feature of quantum systems, and one of the main resources they provide, is entanglement, the correlations between subsystems that cannot be reproduced by classical joint probability distribution. Entanglement underlies many quantum advantages, including teleportation, dense coding, and distributed quantum protocols, and it is the fundamental resource distributed and manipulated by quantum networks. Conversely, it is also one of the features of quantum systems that leads to exponentially high cost of simulation on classical computers, thus necessitating a wide range of bespoke modeling techniques for specific efficient edge cases. Lastly, entanglement degrades due to open system dynamics, thus accurately modeling not only perfect "unitary" evolution but also the decoherence due to interaction with the environment is a requirement for any of these bespoke modeling techniques.

For more complete introductions to these concepts, see standard QIS textbooks and lecture notes, such as Refs.~\cite{nielsen2010quantum, preskill1998lecture}.

\subsection{Stabilizer Formalism and other Restricted Low-complexity Formalisms}

Classical simulation of generic quantum dynamics is intractable in the worst case, so practical simulators rely on identifying (or adaptively exploiting) structure that yields an efficient representation. Fig.~\ref{fig:simmethods} summarizes two complementary sources of efficiency: (i) sparse representations of classical uncertainty and (ii) restricted representations of quantum correlations and entanglement.

\paragraph{Stabilizer formalism.}
For qubit systems, the stabilizer formalism represents a state by the Abelian subgroup of the Pauli group that stabilizes it. Clifford operations (generated by Hadamard, Phase, and CNOT) map Pauli operators to Pauli operators under conjugation, so stabilizer states evolved under Clifford circuits can be updated efficiently using "tableau" methods, and Pauli measurements can be simulated with polynomial-time update rules~\cite{gottesman1998heisenberg, aaronson2004improved}. Many network- and error-correction primitives are close to this regime: ideal Bell-pair generation, entanglement swapping, and syndrome extraction are naturally expressed in terms of Clifford operations and Pauli observables, and common noise models can be approximated or represented as stochastic Pauli channels. Thus supporting this technique is crucial for many of the typical applications for quantum network simulators. QSavory relies on the independent QuantumClifford library for this type of simulations~\cite{quantumclifford}.

\paragraph{Gaussian quantum information.}
For bosonic modes and continuous-variable models, Gaussian quantum information provides an efficient formalism when the multi-mode state is Gaussian and the dynamics preserve Gaussianity. Gaussian states are fully characterized by their first and second moments, and Gaussian operations (generated by Hamiltonians at most quadratic in the canonical operators) act by affine symplectic transformations on these moments. This yields simulations whose cost scales polynomially in the number of modes for large classes of optical and electromechanical network models, including linear optics, squeezing, Gaussian noise channels, and homodyne-type measurements~\cite{weedbrook2012gaussian,bartlett2002efficient}. Quantum networks are almost exclusively optical, and a vast array of optical resources are described by Gaussian states, making these capabilities crucial for a full-stack realistic simulation of quantum networks. QSavory relies on Gabs, another independent simulator library for this type of modeling~\cite{kille2025gabs}.

\paragraph{Tensor networks.}
Restricted formalisms are efficient because the representation is known \emph{a priori}. A more flexible alternative is to \emph{discover} an efficient representation during simulation by compressing the quantum state into a tensor network (e.g., matrix product states/operators, projected entangled-pair states) and adapting its bond dimension as entanglement grows~\cite{schollwock2011density,orus2014practical}. Tensor-network methods can simulate dynamics far beyond Clifford or Gaussian sets, but their cost is controlled by the entanglement structure: they are efficient when entanglement across relevant cuts remains limited or well-structured and become expensive when generic volume-law entanglement develops~\cite{vidal2003efficient}. QSavory currently does not have a tensor network backend, but that project is underway.

\paragraph{Finite-rank stabilizer methods and their relation to tensor networks.}
Between exact stabilizer simulation and fully general exponentially-expensive state vector simulation lie finite-rank methods. A generic state (or channel) can be approximated as a linear combination or quasi-probability mixture of a limited number of stabilizer objects ("stabilizer rank" and related near-Clifford methods)~\cite{bravyi2019simulation}, enabling simulation of circuits with a small amount of non-Clifford "magic" by paying a cost proportional to the chosen rank. Conceptually, this is analogous to tensor-network compression: both approaches trade accuracy for efficiency by restricting the effective dimension of the representation, but they do so in different bases (Pauli/stabilizer structure versus entanglement structure). Similar approaches build not only upon Stabilizer state simulators but also upon Gaussian state simulators. Hybrid strategies exploit whichever notion of low complexity is present in a given model, e.g., near-Clifford dynamics with modest entanglement, or weak non-Gaussianity with locality constraints. The backend simulator libraries we have chosen have these advanced capabilities, but we have not made them available in QSavory yet.

\paragraph{Monte Carlo trajectories}

Of note, even when having to pay the full exponential cost of general-purpose quantum simulations, QSavory avoids a common pitfall in many other simulators. Given that in such a network simulation a common practice will be to run many repeated samples (in a Monte Carlo fashion), we do not use density matrices, instead we use Monte Carlo over quantum pure trajectories. All noise processes (e.g. Lindbladian dynamics) and measurements are properly handled nonetheless, leading to the exact same overall statistics of the results. This leads to reduction of simulation cost from $4^n$ to $2^n$ where $n$ is the number of entangled qubits, because we model densely only the quantum amplitudes, not the classical probability distributions, i.e. we stay in the upper right corner of Fig.~\ref{fig:simmethods}.






\subsection{Quantum Networks}

\begin{figure} [tb]
        \centering
        (a) Two-way
        {\includegraphics[width=\linewidth]{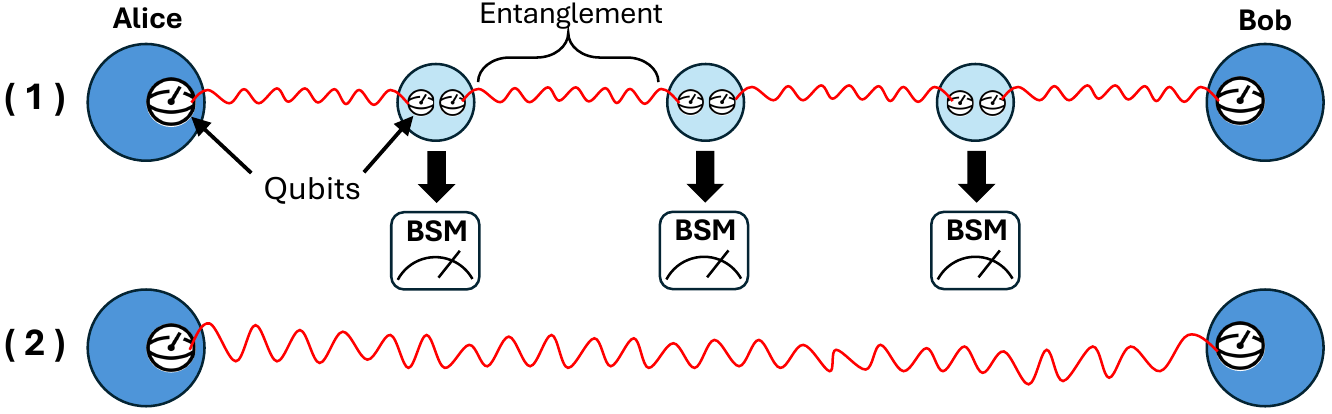}}
        
        (b) One-way
        {\includegraphics[width=\linewidth]{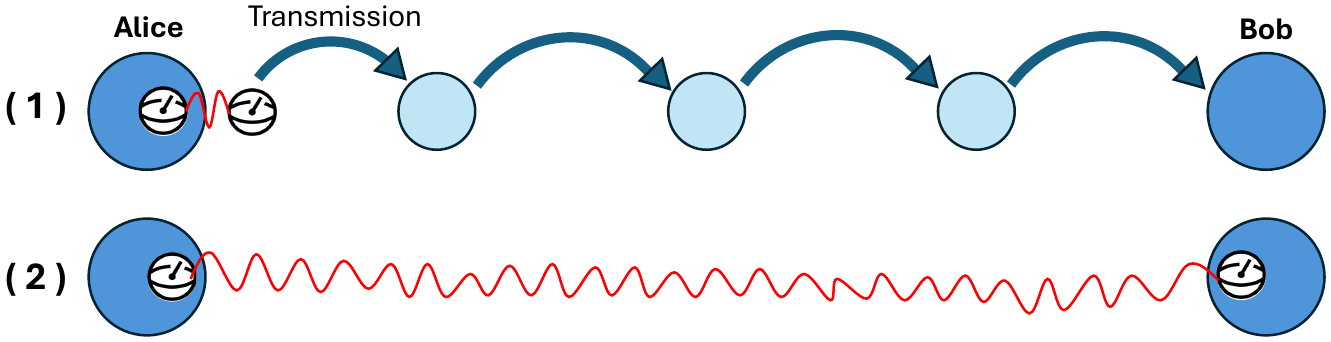}}
        \caption{The operation of (a) two-way and (b) one-way quantum networks to distribute entanglement. End nodes and repeaters are in blue. Red lines represent an established Bell pair. "BSM" (Bell state measurement) is a process which can enable an "entanglement swap", turning two short-distance pairs into one long distance pair.}
        \label{fig:one_two_way}
\end{figure}

\begin{figure} [tb]
        \centering
        \hspace{0.5cm} (a) Distillation \hspace{1.8cm}  (b) Error Correction
        \vspace{0.25cm}
        
        \includegraphics[height=3.3cm]{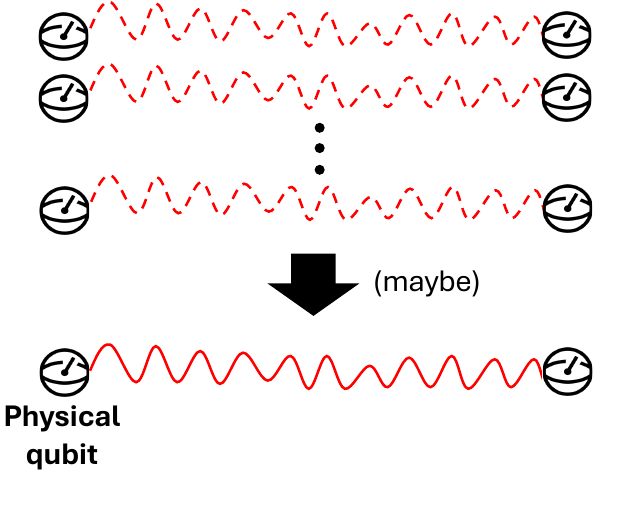}
        \includegraphics[height=3.3cm]{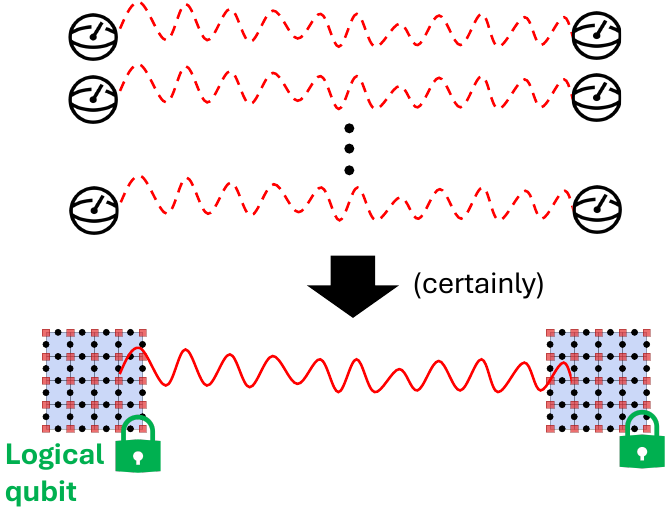}
        \caption{A high-level illustration of (a) entanglement distillation and (b) quantum error correction. Dashed lines represent low-quality entanglement, solid lines signify improved quality (a.k.a.\ fidelity). While the error-correction approach is not heralded, i.e.\ it always reports a success, it might suffer from more "false positives". The probabilistic distillation process offers a different tradeoff between rate of entanglement and fidelity, a central question studied in quantum networking.}
        \label{fig:dist_corr}
\end{figure}

A quantum network distributes entangled states (e.g., Bell pairs) or directly transfers quantum states among remote users, providing the basic resource for distributed quantum applications. Entanglement is typically generated between directly connected nodes and extended across multiple hops until it reaches the desired endpoints. Designing such networks requires addressing two fundamental problems:
\begin{itemize}
    \item the impossibility of amplifying quantum signals due to the no-cloning theorem, which forces alternative methods to counteract the exponential loss of photons over optical fibers;
    \item the presence of quantum errors and memory decoherence, which limit the number of operations and the duration for which quantum states can be reliably stored.
\end{itemize}

As illustrated in Fig.~\ref{fig:one_two_way}, a convenient classification of quantum networks can be done first according to their operating principle.  
In \emph{two-way} networks (Fig.~\ref{fig:one_two_way}a), neighboring nodes repeatedly attempt to establish pre-defined entangled states until every link along a path connecting the end users has generated entanglement. Intermediate nodes then perform "entanglement swaps" through local Bell-state measurements (BSMs) that take two "shorter" entangled pairs and create a longer pair across the entire path~\footnote{This ordering can be less restrictive as every intermediate node can, in principle, perform the BSM as soon as the two entangled states that are local to it are generated}. This probabilistically-generated long-range entanglement resource can then be used to teleport an arbitrary qubit from one node to another or to enable other QIS protocols. The intermediate nodes are commonly referred to as \emph{quantum repeaters}.

In contrast, \emph{one-way} networks (Fig.~\ref{fig:one_two_way}b) forward arbitrary quantum states directly through a sequence of connected nodes. A qubit prepared at the source (e.g.\ one half of an entangled pair) is physically transmitted over quantum channels and processed hop-by-hop until it reaches the destination. This paradigm enables packet-style routing of quantum information, conceptually similar to classical forwarding, although it is possible to have packet-forwarding abstractions in two-way networks as well \cite{Bacciottini_2025}. Somewhat tautologically, the physically transmitted qubit can be used to directly distribute entanglement (thus mimicking the base resource prepared in two-way networks), or it can be used directly for specific QIS protocols.

Beyond this operational distinction, quantum networks can also be classified by how they manage noise and transmission errors (Fig.~\ref{fig:dist_corr}).  
\emph{Entanglement distillation} protocols consume several noisy entangled states (dashed lines in Fig.~\ref{fig:dist_corr}a) to probabilistically produce a smaller number of high-quality entangled states.  
\emph{Quantum error correction}, instead, deterministically encodes a single logical qubit into multiple physical qubits, yielding an entangled state that is both of high quality and protected from future errors (Fig.~\ref{fig:dist_corr}b).  
Error correction becomes advantageous once the underlying hardware achieves gate fidelities above the fault-tolerance threshold, whereas distillation is often favored in near-term, noisy platforms.

Previous works have explored architectures encompassing nearly every combination of the above design axes~\cite{Munro_2014, Mantri_2025, Van_Meter_2022, DiAdamo_2022, Muralidharan_2014}---for instance, two-way networks with distillation, or one-way networks with error correction. Further diversity arises from assumptions about the underlying hardware. This variability highlights the need for a simulator that can model heterogeneous architectures. In such a simulator, modularity is essential: updating the physical model of a node should not require rewriting higher-level logic, such as error correction or entanglement management. The capability to model more than just qubits is also crucial, as the richness of optical communication is difficult to study with qubit state-vector models: either because error correction requires hundreds of qubits and specialized fast simulation techniques, or because the natural physical model for optical modes involves bosonic quantum states, not qubits.

\section{QuantumSavory\label{qsavory}}
\begin{figure} [tb]
        \centering
        {\includegraphics[width=0.75\linewidth]{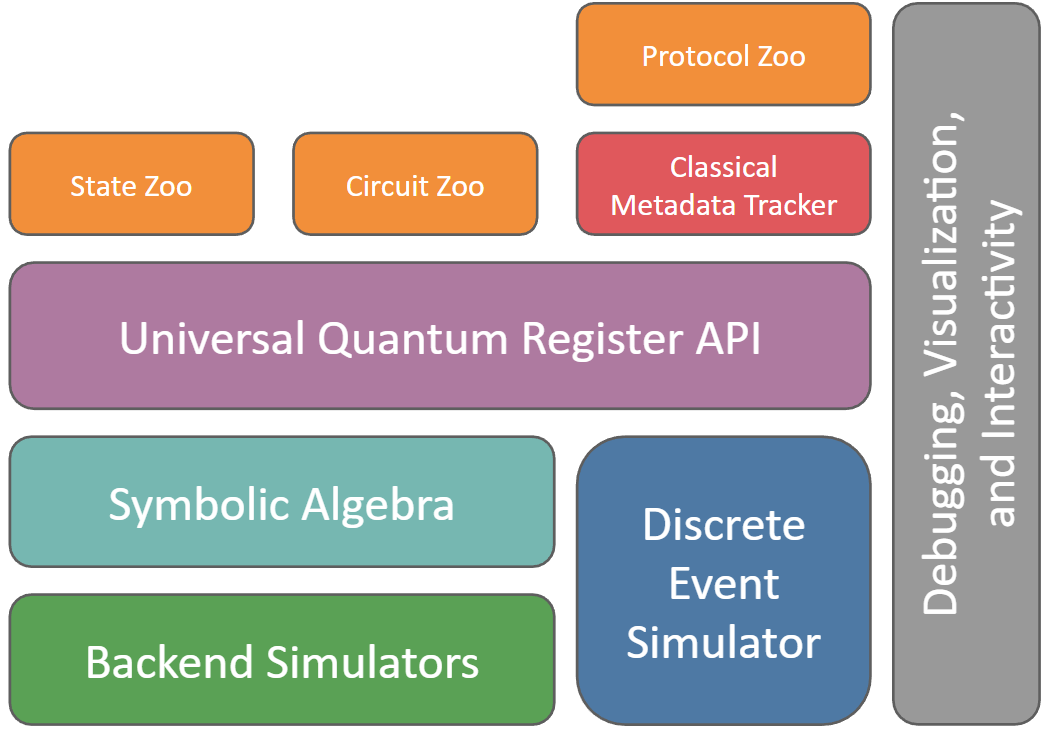}}
        \caption{High-level architecture of QSavory, centered on the universal quantum register API that connects modular \textit{Zoos} (states, circuits, and protocols) with backend-agnostic symbolic representations, interchangeable simulators, and a discrete-event classical control layer supporting LOCC workflows, with debugging and visualization spanning the full stack.}
        \label{fig:qsavoryblocks}
\end{figure}
This section introduces the design and core abstraction of QSavory. We begin with an end-to-end example, illustrating how a graph state generation protocol can be expressed in a few lines of code. We then describe the primitives that enable this workflow: an API for controlling quantum registers with declarative configuration of properties such as the type of the physical system or the background noise processes they experience; symbolic backend-agnostic state descriptions; discrete-event-based classical process control for LOCC protocols; a tag/query messaging layer enabling lightweight and easily composable coordination between classical control components; and modular \textit{Zoos}, i.e.\ libraries of commonly used quantum states, circuits, and full-blown protocol building blocks. Fig.~\ref{fig:qsavoryblocks} summarizes how these components fit together in the overall QSavory architecture.

\subsection{Overview}\label{subsec:overview}
A complete quantum network simulation in QSavory can be set up in just a few lines of code. The goal of this overview is not to explain every function call in detail---that will come in the following subsections---but rather to draft the overall structure of a simulation and how the different components fit together. Even if some of the syntax may appear unfamiliar, the reader should keep in mind that the main focus here is on the semantics rather than the implementation details.

\begin{figure} [tb]
        \centering
        {\includegraphics[width=\linewidth]{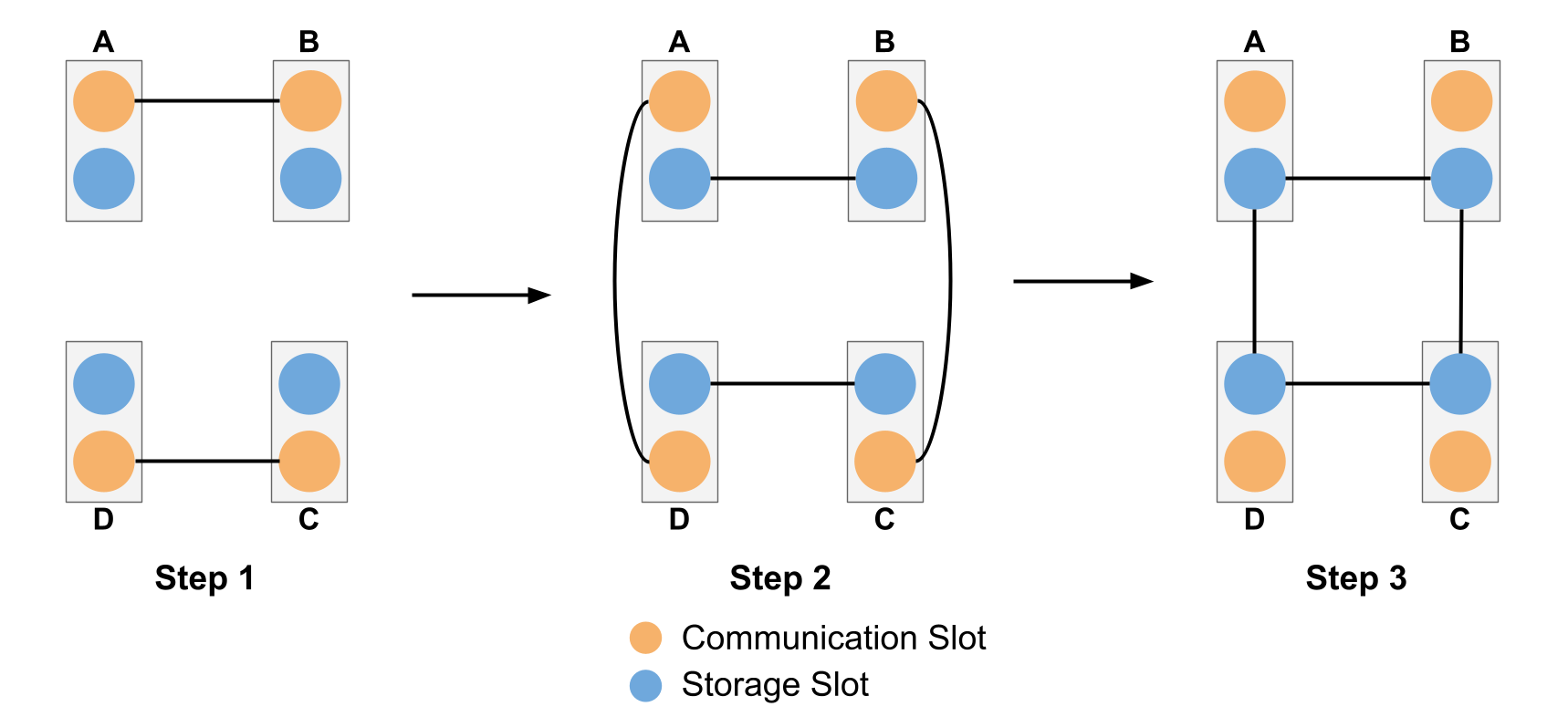}}

        \caption{The cluster state preparation example from the QSavory overview section. \texttt{EntanglerProt} is called on opposing edge pairs to entangle communication qubits. Two opposite edges are attempted first since they do not overlap and can be done in parallel. After the first round of entanglement generation, the quantum states are transferred to the storage slots. Then, the remaining edges are entangled (again through their communication qubits). Lastly, fusion circuits are executed, fusing the four pairs into a single cluster state stored in the storage qubits.}
        \label{fig:overview}
\end{figure}

We simulate the distribution of a four-qubit cluster state across four network nodes A-D arranged in a square, described in Fig.~\ref{fig:overview}. Each node has two qubits, a communication qubit that has the capability to be entangled with another node, and a storage qubit where entanglement is stored long term. This example will use a standard circuit for entanglement "fusion", linking already entangled storage nodes to other storage nodes (through the consumption of the communication node entanglement) into larger entangled states. The example will showcase the quantum Register control API abstraction, the use of pre-defined circuits and control protocols, the discrete event simulator permitting parallel simulated processes, and the independence of the backend quantum dynamics simulator from the symbolic frontend.

QSavory represents each node with a \texttt{Register}, which encapsulates local quantum subsystems (e.g., qubits, qmodes) together with their noise models at the desired level of abstraction. The four registers are collected in a \texttt{RegisterNet} for coordination; background noise processes (e.g., \texttt{T1Decay}) can be declared at construction time.

\begin{lstlisting}[language=Julia]
# slot 1 is the communication slot, and slot 2 is storage
comm_slot, storage_slot = 1, 2

# a register with 2 qubits, both suffering T1 noise
A = Register(2, T1Decay(1.0))
# [...] Do the same for B,C,D.

net = RegisterNet(cycle_graph(4), [A, B, C, D])
sim = get_time_tracker(net)
\end{lstlisting}

The storage qubits are first initialized in the first eigenstate of the $X$ operator (the $|+\rangle$ state) using \texttt{initialize!}, a standard choice for typical entanglement fusion circuits:
\begin{lstlisting}[language=Julia]
for reg in [A, B, C, D]
  # initialize to the first eigenstate of the X operator
  initialize!(reg[storage_slot], X1)
end
\end{lstlisting}

Afterwards, we need to start entangling neighbors. The type of entanglement we want (a Bell state) can be expressed as the state stabilized by the operators $Z\otimes X$ and $X\otimes Z$. While this is a stabilizer-state notation, it will be translated to any backend we are using (e.g.\ state vector or Clifford formalism or another numerical method). The next step is to set up the distribution of these Bell pairs. Instead of doing this manually, we will showcase the use of predefined frequently-used protocols, many of which are available in QSavory's \textit{Zoos} -- curated libraries of off-the-shelf circuits, states, and protocols. For instance, we can directly instantiate a generic entanglement-generation protocol (\texttt{EntanglerProt}), or parameterize it into a more specialized one without altering the rest of the simulation. The protocol results in (i) initializing the qubits in the desired entangled state, including any effects of hardware imperfections or delays due to communication or synchronization, and (ii) "tagging" the entangled register slots with meta data 
that allows other protocols to identify and wait for the presence of an entanglement resource without having a direct handle to the entangling protocol. In the snippet below, we launch independent entanglers between all node pairs:

\begin{lstlisting}[language=Julia]
# Use an off-the-shelf entangling protocol:
entangler_AB = EntanglerProt(sim, net, 1, 2,
                 chooseslotA=comm_slot, chooseslotB=comm_slot,
                 pairstate=StabilizerState("ZX XZ"))
# Register the protocol as one of
# the parallel processes in the simulation:
@process entangler_AB()
# [...] Do the same for BC,CD,DA. 
\end{lstlisting}

"Protocols" defined by the user or available in the Protocol Zoo can also rely on locks, wait on events, or wait for changes in the local classical metadata, in order to coordinate resource use and avoid conflicts. In this example, only the communication qubit can be entangled with that of the other nodes, yet it must participate in two pairwise entangling operations. While the qubit is in use, all other processes must wait for it to be freed. QSavory’s resource management ensures that pairs are generated and consumed sequentially without conflicts, and the user only needs to specify what should happen rather than how to manage concurrency.

For the next step, instead of reusing a pre-defined protocol we will create a custom one. Once the Bell pairs are available in the communication qubits, they need to be moved to the storage slots. QSavory's \textit{Zoo}s implement a standard \texttt{Fusion} circuit that we can use in our custom protocol. The user-defined \texttt{fusion} protocol below expresses this logic succinctly: whenever a Bell pair qubit is tagged, the corresponding fusion circuit is executed.

\begin{lstlisting}[language=Julia]
@resumable function fusion_protocol(sim, node)
  while true
    # Wait for and consume one available Bell-pair tag.
    entry = @yield querydelete_wait!(
      node, EntanglementCounterpart, W, W, W;
      assigned=true)

    _, remote_node, _, _ = entry.tag

    # ... also delete the reciprocal EntanglementCounterpart to keep things neat

    circuit = Fusion()
    circuit(node, net[remote_node], comm_slot, storage_slot)
  end
end
\end{lstlisting}

Only nodes A and C need to install this (user-defined) fusion protocol, since each of them sits at the junction of two edges in the cluster square:
\begin{lstlisting}
@process fusion_protocol(sim, A)
@process fusion_protocol(sim, C)
\end{lstlisting}

The distinction between a "protocol" and a "circuit" is that the protocol might involve communication, waiting for events, and overall a "discrete event simulation", while a circuit is just a sequence of gates and measurements (without waiting for communication from other nodes or other asynchronous events).

This simulation can, with minimal changes, incorporate new noise models, larger networks, distillation, or alternative entanglement-generation schemes, supporting, for example, performance evaluation of complex distributed protocols from a networking perspective or the accurate simulation of a quantum hardware platform such as color centers.

In the subsections that follow, we will take a closer look at the main components and functionalities that appeared here---registers, quantum dynamics, classical metadata tags, classical communications, protocols, and predefined \textit{Zoo}s of such resources.

\subsection{Quantum Modeling}\label{subsec:modeling}
One of the most defining features of QSavory is its ability to model various quantum systems in different formalisms while providing a single frontend. While many simulators are confined to qubits, realistic physical platforms require much more: bosonic modes are natural for optical quantum communication, higher-dimensional qudits represent more realistic hardware, and hybrid architectures couple different types of subsystems. QSavory’s register abstraction was created to make this kind of modeling natural. A single register can hold qubits or harmonic oscillators, each with its own background noise processes. This allows composite nodes to be expressed directly, without stitching together disparate tools. Moreover, different numerical formalisms are available for more efficient modeling depending on the type of the subsystem, e.g., the stabilizer formalism for qubits or the Gaussian formalism for bosonic modes.

The register interface in QSavory provides users with a compact yet expressive way to specify and manipulate quantum systems. States can be initialized symbolically and are automatically translated into the numerical representation required by the chosen backend. Operations may target one or several subsystems at a time, and the simulator only composes Hilbert spaces when those subsystems actually interact. This means that states are stored in a fashion that keeps them as factored out as possible: instead of forming a single large joint state vector or density matrix, QSavory maintains a collection of smaller states whenever no entangling operations have occurred. Because memory grows exponentially with the number of qubits (in the state vector formalism), preserving this factorization dramatically reduces memory consumption and enables state vector simulations of large, structurally sparse quantum systems that would otherwise be intractable. Observables, expectation values, projective measurements, and partial traces operate directly on these factorized Hilbert spaces and automatically merge or reduce them only when necessary.

A second, complementary feature is QSavory's approach to time evolution and noise. Background processes such as decay, dephasing, and depolarization are specified declaratively at the moment the register is constructed: users state which subsystems experience which noise models, without having to manually apply these processes throughout the simulation. The evolution induced by these processes is then carried out \emph{lazily}: each subsystem maintains its own local simulation clock and advances only when demanded by a user-level operation. Different parts of the same entangled system may therefore evolve at different rates until an operation forces them to synchronize. This separation between declarative noise specification and demand-driven evolution ensures that the simulator expends computational effort only when necessary, avoiding unnecessary updates and grouping together as many operations as possible, thus reducing overhead. Moreover, the user does not need to manually compose the operations they want to perform (gates, measurements, even time-dependent Hamiltonians) with the background noise processes. The necessary master equation is derived on the fly by the backend.

Another strength of the framework is its symbolic, formalism-agnostic front end~\cite{kille2025qsymbolics}. States and operations are written in a symbolic language and translated to a specific numerical backend library that reflects the underlying quantum representation. In QSavory, different numerical representations can be seamlessly integrated \footnote{provided that well-defined interface functions for symbolic conversions are invoked on data structures in the backend library}.

Building on this flexibility, QSavory already supports multiple numerical backends for well-established formalisms of quantum simulation. For example, a model can run using QuantumClifford.jl~\cite{quantumclifford} for Clifford simulations, QuantumOptics.jl~\cite{kramer2018quantumoptics} for full state-vector evolutions, or Gabs.jl~\cite{kille2025gabs} for Gaussian phase-space dynamics. Switching between simulation backends is trivial and different parts of a network can use different formalisms at the same time. This flexibility is important in practice: Clifford- or Gaussian-based simulation workloads benefit from polynomial-time performance, while more general operations require full wavefunction methods that scale exponentially. Switching between them is a matter of choosing a backend rather than rebuilding a model from scratch. We emphasize that these are not the only possible backends. The backend simulators are independently developed, and hooked in through a small well-defined API, enabling the future addition of other backends, even directly by the user. As a proof of the ease with which a new backend can be attached, the entire Gaussian-backend integration required fewer than 40 lines of interface code in QuantumSavory.jl (a further ~130 lines provide optional pretty-printing and show methods).

To see these ideas in action, consider a simple transduction / back-action example. Imagine two registers, one for qubit states, and another one for quantum modes. Similarly to how measuring one of two entangled qubits leads to known back-action to the unmeasured qubit, which then can be used to perform teleportation, we can have back-action due to measuring a quantum mode entangled to a qubit. The specific application (transduction or teleportation) depends on the specifics of the initial states and interaction, but here we will simply showcase interesting interaction (e.g.\ Jaynes-Cummings) followed by a measurement type not possible for qubits (a Homodyne measurement):
\begin{lstlisting}[language=Julia, mathescape=true]
register = Register([Qubit(), Qumode()])
# Z1 is the first eigen state of Z
# F0 is the Fock state with zero excitations (vacuum)
initialize!(register[1:2], tensor(Z1,F0))

# entangle the two systems through Jaynes-Cummings interaction
interaction = tensor(Pm, Create) + tensor(Pp, Destroy)
quarter_period = exp(-im * (pi / 4) * interaction)
apply!(register[1:2], quarter_period)

# measure the quantum mode,
# and cancel the backaction on the qubit
q = project_traceout!(register[2], HomodyneMeasurement([0.0]))
correction = exp(im * atan(q) * X)
apply!(register[1], correction)
# and now the qubit is in a known state
\end{lstlisting}
The simulator composes the \texttt{Qumode} and \texttt{Qubit} subsystems only when they interact, ensuring that memory grows with the size of entangled clusters rather than with the full product space. Background noise processes evolve automatically, triggered only when observables are evaluated or gates applied. The entire flow remains symbolic and backend-agnostic until it is expressed in the numerical form needed for the chosen simulator.

\subsection{Modeling Discrete Event Dynamics}\label{subsec:discrete}
Modeling complex distributed systems often requires explicit support for message exchange between entities, synchronization mechanisms, and dynamic system evolution (e.g., users or components joining and leaving). In the quantum-information setting, these requirements often translate into simulating multi-step \emph{local operation and classical communication} (LOCC) protocols, where the timing and structure of local quantum operations are dictated by the classical messages exchanged among the involved parties.

QSavory relies on \emph{discrete-event simulation}, implemented through ConcurrentSim.jl~\footnote{\url{https://github.com/JuliaDynamics/ConcurrentSim.jl}, last accessed: June 2026}, to support message-passing, synchronization, and event-driven interactions between simulated entities. In what follows, we formalize this simulation model and introduce the main simulation pattern adopted throughout QSavory.

In Julia, marking a function with \texttt{@resumable} makes it act as a co-routine, a.k.a.\ a generator, a.k.a.\ a continuation,~\footnote{Generators or co-routines or continuations are a standard concept in many programming languages that provide the user with a handle to a computation that has been cooperatively paused and is waiting to continue. They are frequently employed in the design of discrete event simulators.}, whose execution can be suspended and later resumed each time it \texttt{@yield}s a value.

Within ConcurrentSim.jl, resumable functions model processes that suspend their execution until specific conditions are met---such as the arrival of a message, a timeout, or the availability of a resource. For example, consider a process that waits for a swap request before performing an entanglement swap between two of its register slots:
\begin{lstlisting}[language=Julia]
# a process that waits for a message and then swaps
@resumable function swapper(net, node)
    mb = messagebuffer(net, node)
    
    # Wait for and consume a swap request.
    msg = @yield querydelete_wait!(mb, :swap_request)

    # [...] Perform the local operations for entanglement swap.
end

@process swapper(net, 1) # launches the process on node=1
\end{lstlisting}

 As a convention that makes configuration and reuse of protocols easier, QSavory extends this model by introducing the notion of an \texttt{AbstractProtocol}: a resumable function equipped with a context that bundles the protocol’s runtime information. A user is free to simply use \texttt{@resumable} functions, but "protocols" provide a more standardized format for data storage, visualizations, and logging, and many such protocols are provided in the "Protocol Zoo". The configuration context of a protocol typically includes references to the simulation and the \texttt{RegisterNet}, and may also contain parameters such as the node(s) on which the protocol runs or additional configuration settings. The following example illustrates how a swapper can be implemented using this style:
 \begin{lstlisting}[language=Julia]
struct MySwapperProt <: AbstractProtocol
    sim::Simulation
    net::RegisterNet
    node::Int # where the swapper runs
    alice::Int
    charlie::Int
    # Add optional configuration fields here.
end

# We make the protocol callable as a resumable function
@resumable function (prot::MySwapperProt)()
    (;sim, net, node, alice, charlie) = prot
    
    # [...] same code as before
end

# usage:
my_swapper = MySwapperProt(sim, net, 2, 1, 3)
@process my_swapper()  # starts the protocol on node=2
\end{lstlisting}

Protocols can suspend execution while waiting for a variety of conditions, which typically include: (i) waiting for a specified delay via \texttt{timeout(sim, delay)}; (ii) waiting for a tag (see next section) to appear on a register, such as \texttt{onchange(reg)} or \texttt{query\_wait(reg, MyTagType, ...)}; (iii) waiting for the arrival of a message in a message buffer, e.g., \texttt{onchange(mb)} or \texttt{query\_wait(mb, MyMsgType, ...)}; and (iv) synchronizing on shared resources, such as acquiring a lock on a register slot with \texttt{lock(reg[slot])}.
These mechanisms form the basis for defining event-driven control flow within QSavory protocols.

An additional useful feature is that conditions can be combined through logical operators:

\begin{lstlisting}[language=Julia]
@yield (lock(q1) & lock(q2)) # both slots must be locked

# waits at most 10.0 time units for a message on mb
@yield (onchange(mb) | timeout(sim, 10.0))
\end{lstlisting}

\subsection{Tags, Queries, and Messaging}\label{subsec:tags}
   Simulators of composite systems must support interactions among connected sub-entities, such as two hosts attached to a network running a variety of different protocols that compete for resources. In practice, this is typically achieved either by explicitly wiring entities together and enabling message passing (by giving each entity a handle to the other entities or by explicitly constructing message channels between entities), or by exposing a shared state abstraction that entities can read and update. We covered quantum-specific interactions in Section~\ref{subsec:discrete}; here we instead focus on how QSavory handles \emph{classical} interactions, including communication and synchronization tasks. While existing quantum simulators follow the explicit-connection approach, QSavory adopts a shared messaging queue abstraction. This design choice is beneficial in many networking settings and employed by well-established classical queuing and networking systems like ZeroMQ, RabbitMQ, and more. As we show in the rest of this section this pattern streamlines both the tracking of quantum states and the orchestration of distributed operations within the simulation.

The tagging and querying infrastructure provides a uniform way to attach metadata to different entities in a simulation and later retrieve it through declarative queries. Conceptually, it turns the simulator into a lightweight distributed database where protocols can store, search, and act on information without needing to know how or when it was produced or who produced it. The basic interface works as follows:
\begin{lstlisting}[language=Julia]
# Imagine a tag as a custom list of labels
# that can be symbols, numbers, or simple data structures
tag!(entity, tag) # attach tag to entity
query(entity, tag) # search for any matching element
queryall(entity, tag) # search for all matching elements
querydelete!(entity, tag) # deletes after successful search
\end{lstlisting}
Tags in QSavory can be rich data structures and can be attached to any queryable entity, most prominently \texttt{Register} slots and the \texttt{MessageBuffer}. For registers, tags provide a structured way to manage metadata about quantum resources on a network. They allow, for example, entanglement to be tracked by tagging a register slot with the information about its entanglement counterpart, ensuring that later processes can locate the correct partners without manual bookkeeping and without having to possess a handle or a direct messaging channel to the process that created the entanglement.
\begin{lstlisting}[language=Julia]
# EntanglementCounterpart is a standard ProtocolZoo tag.
# Its fields identify the remote node, remote slot, and pair.

# Tag the register slots
# (nodeA is node 1 and nodeB is node 2 of the network)
tag!(nodeA[1], EntanglementCounterpart, 2, 1, 101)
tag!(nodeB[1], EntanglementCounterpart, 1, 1, 101)

# Later, find the slots by tag.
# You can use W as a wildcard for any field.
# E.g. a query for a slot tagged with
# EntanglementCounterpart (to nodeB with any slot and id):
res = query(nodeA, EntanglementCounterpart, 2, W, W)
\end{lstlisting}

The main value-add provided by this tagging system is that protocols can now be composed simply by agreeing on a set of basic tags (e.g.\ presence and quality of entanglement, history of performing a swap, purification, or error correction, fusing into a larger state, etc), instead of by requiring complex dependencies, manual linking through explicit handles or messaging channels, etc. This simple idea turns many different technical problems into a much simpler social problem: just agree on the meaning of a tag. Compare this to other modeling software architectures where you either (1) have access only to much simpler resources, like exclusively Bell pairs of qubits; (2) have to obey a strict object oriented inheritance structure (with the well-known drawback that inheritance has compared to composition); or (3) protocols working cooperatively have to have explicit handles for each other or manually created message channels.

Moreover, the querying system is able to search for tags not only by exact match, but also through wildcards for certain fields (i.e.\ matching any value for that field), or even by arbitrary predicates (a condition that the value needs to meet), mimicking the capabilities of database query languages like SQL.

Besides storing information at a slot, the tagging and querying infrastructure can naturally be used for passing classical messages between nodes. Message buffers collect these classical messages arriving at a network node. In a buffer, each tag corresponds to an incoming message, and the simulator automatically manages their insertion as communication events occur. This abstraction frees protocol designers from explicitly implementing message handling, letting them focus on higher-level protocol logic.
A key advantage is that multiple protocols can share the same buffer while remaining unaware of each other. Each protocol simply queries for the tags relevant to its operation, effectively subscribing to its own "topic". This mirrors the design of message queuing systems in cloud architectures (e.g., RabbitMQ~\footnote{\url{https://www.rabbitmq.com}} or Kafka~\footnote{\url{https://kafka.apache.org/intro}}), where producers and consumers are decoupled. Just as this model has improved the scalability of modern microservice architectures, QSavory brings the same paradigm to quantum network simulation: processes can be composed declaratively by agreeing on tag labels rather than hard-wired interfaces.

Buffers also naturally integrate with register tags. For example, an incoming message may herald the success of an entanglement distillation round; in response, the protocol can attach a tag to the corresponding qubit in a register, marking it as distilled and ready for use in higher-level protocols. 
To illustrate, we complete the swapper example introduced in Section~\ref{subsec:discrete} by explicitly describing the wait condition query:

\begin{lstlisting}[language=Julia]
# Somewhere else: initialize Bell pairs...
initialize!(...) #Alice-Bob pair
initialize!(...) #Bob-Charlie pair

# ...and tag the slots held by Bob
tag!(..., EntanglementCounterpart(...)) # with Alice
tag!(..., EntanglementCounterpart(...)) # with Charlie

@resumable function (prot::MySwapperProt)()
    (;sim, net, node, alice, charlie) = prot
    mb = messagebuffer(net, node)
    reg = net[node]

    # Wait for a swap request, consuming it on arrival
    msg = @yield querydelete_wait!(mb, :swap_request)

    # Now the actual swap operation:
    # find a slot entangled with Alice and one with Charlie
    a = query(reg, EntanglementCounterpart, alice, W, W)
    b = query(reg, EntanglementCounterpart, charlie, W, W)

    # [...] lock and revalidate both slots, then delete their local tags

    # Local Bell-state measurement (returns two Pauli eigenvalues, +/-1)
    circuit = LocalEntanglementSwap()
    xmeas, zmeas = circuit(a.slot, b.slot)

    # [...] send endpoint updates through classical channels and release the locks

end

@process MySwapperProt(sim, net, 2, 1, 3)()
\end{lstlisting}

In this example, the swapper node does not need to know who sent the request, only that a \texttt{:swap\_request} tag appeared. Even more importantly, the entities requesting the swap do not need to own a reference (handle) to the swapper protocol, providing a level of modularity and composability difficult to achieve with other software systems.

Underlying this infrastructure are \textbf{classical links}, which define the topology of the classical communication network. A channel between adjacent nodes uses their direct link. For nonadjacent nodes, the user can opt into shortest-path forwarding with \texttt{permit\_forward=true}; the simulator then forwards the message through intermediate nodes. Propagation delays are handled within the discrete-event engine, so communication latencies can reflect physical constraints such as speed-of-light delays. We remark that this API models classical message passing between registers, and there is a separate \texttt{QuantumChannel} API to model explicit transmission of quantum states, if needed (as an alternative to the more abstract  \texttt{EntanglerProt} used to simulate many different standard entangling methods).

On arrival, messages are placed in the destination node's \texttt{MessageBuffer}, where they become immediately available through querying (or a query might even have already paused the protocol, waiting to continue once a message is received). To complete the previous example, a user would request a swap with:
\begin{lstlisting}[language=Julia]
put!(
    channel(net, node => swapper_node; permit_forward=true),
    :swap_request,
)
\end{lstlisting}

To conclude, we recall that the tags-and-queries mechanism not only enables communication between protocols but also provides a uniform way to define the inputs and outputs of each protocol through well-typed messages and tags. This uniformity is what makes it possible to assemble a \texttt{ProtocolZoo}: a collection of interoperable building blocks that can be combined or seamlessly replaced across different simulation scenarios.
The next section provides an overview of the "Zoos" shipped with QSavory.

\subsection{Zoos: Modular Repositories for Simulation}
QSavory introduces a modular architecture that makes it easy to build with reusable components, organized into compendiums referred to as \textit{Zoos}: the StatesZoo, CircuitZoo, and ProtocolZoo. Instead of repeatedly constructing states, circuits, or protocols from scratch, users can directly employ off-the-shelf modules without needing to know the technical details of the implementation.

\subsubsection{StatesZoo}

\begin{figure}
        \centering
        \includegraphics[width=\linewidth]{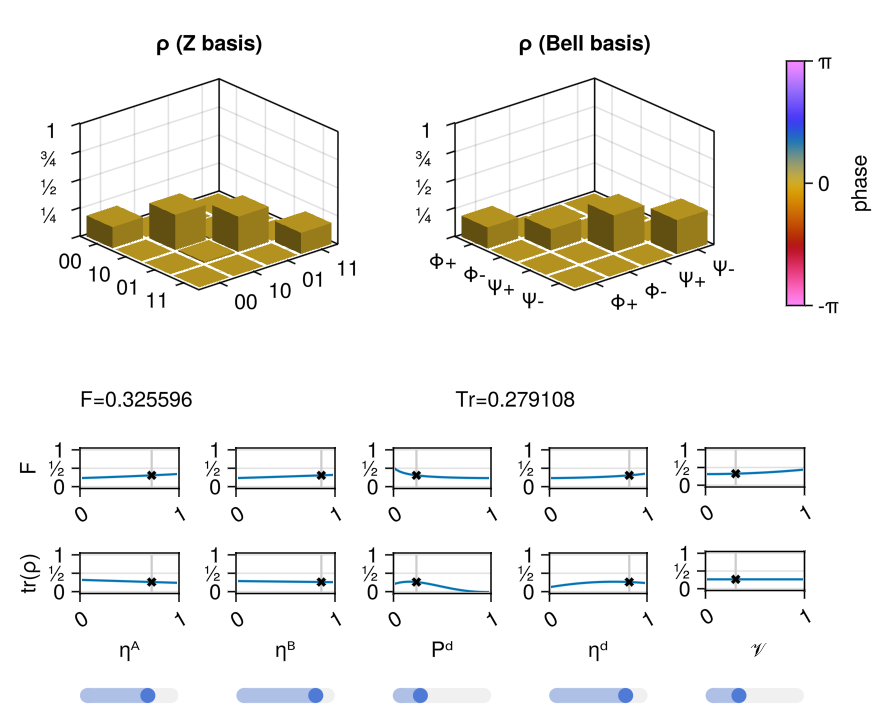}
        \caption{The "State explorer", a tool used to visualize entries in the StatesZoo, showing the Barrett-Kok entangled pair. The sliders let users modify various parameters of the model of the hardware producing the entangled pair. Various figures of merit are plotted, and each inset shows how these figures of merit would change when sweeping a given parameter.}
        \label{fig:bkslider}
\end{figure}

The StatesZoo is a curated library of highly parameterizable symbolic representations of commonly used quantum states. It enables users to instantiate realistic states produced by complex physical processes without constructing the underlying low-level dynamics from scratch. For example, it contains physically accurate models for noisy entanglement sources like ZALM~\cite{zalm, prajit2022heralded, richardson2025full} or the Barrett-Kok procedure~\cite{barrett-kok}. By specifying the relevant physical and experimental parameters of the  procedure, users can directly initialize a pair of qubits in the entangled state predicted to be produced.
\begin{lstlisting}[language=Julia]
reg = Register(2)
initialize!(
  reg[1:2], BarrettKokBellPair(etaA, etaB, dark_counts, ...))
\end{lstlisting}

QSavory also provides a lightweight “state explorer” for interactively visualizing predefined black-box states from the StatesZoo and exploring how their properties depend on model parameters. The state explorer for the Barrett-Kok pair is shown in Fig.~\ref{fig:bkslider}. Users can select a surrogate model for the physical system of interest, vary its parameters through interactive controls, and immediately observe the resulting changes in relevant quality metrics, without needing to inspect or modify the underlying implementation.

We are continuing to expand the StatesZoo with additional surrogate models representing a broader range of experimentally relevant quantum systems.

\subsubsection{CircuitZoo}

The CircuitZoo provides circuits that are common in quantum communication and networking contexts, such as entanglement swaps and fusion, superdense coding, and more. Notably, CircuitZoo includes a range of entanglement purification circuits, from a simple 2-to-1 purification routine to advanced circuits for specific noise models~\cite{krastanov2019optimized, nickerson2013topological}.

The following example demonstrates how the "double selection" 3-to-1 entanglement purification circuit~\cite{fujii2009double} can be called succinctly in QSavory. The circuit, \texttt{Purify3to1}, is parameterized by two symbols specifying which Pauli error each constituent purification subcircuit leaves undetected. With distinct choices such as \texttt{:Z} and \texttt{:Y}, their checks together cover all three Pauli error types.

\begin{lstlisting}[language=Julia]
a = Register(3)
b = Register(3)
bell = StabilizerState("ZZ XX")
for i in 1:3 initialize!((a[i], b[i]), bell) end

# purifiedL, purifiedR, followed by 
# sacrificedL1, sacrificedL2, sacrificedR1, sacrificedR2
Purify3to1(:Z, :Y)(a[1], b[1], a[2], a[3], b[2], b[3])
\end{lstlisting}
Here, the second and third Bell pairs are sacrificed to purify the first. If an error is detected, the circuit returns \texttt{false} and the state is reset. If no error was detected, the circuit returns \texttt{true}.

Similar to the StatesZoo, the CircuitZoo is under active development; recent additions include a fusion operation for protocols based on graph states.

\subsubsection{ProtocolZoo}
The ProtocolZoo complements the StatesZoo and CircuitZoo by providing ready-to-use, composable protocol modules for tasks such as entanglement generation, swapping, routing, and the management of classical control metadata across a network.

For example, a complete repeater-chain workflow can be assembled entirely from existing modules: \texttt{EntanglerProt} generates entangled pairs between neighboring nodes and tags the corresponding qubits with \texttt{EntanglementCounterpart}; \texttt{SwapperProt} then performs entanglement swapping, while the optional \texttt{CutoffProt} periodically discards stale qubits whose retention time has expired; meanwhile, \texttt{EntanglementTracker} processes coordination messages to keep local metadata consistent with changes at remote nodes.

In QSavory, protocols are implemented as resumable functions, also known as generators or coroutines or continuations. This allows multiple protocol instances to execute concurrently within the discrete-event simulation. Each protocol is encapsulated in a structure containing the configuration parameters. For example, \texttt{EntanglerProt} is defined as follows:

\begin{lstlisting}[language=Julia]
@kwdef struct EntanglerProt <: AbstractProtocol
    sim::Simulation
    "a network graph of registers"
    net::RegisterNet
    "the vertex index of node A"
    nodeA::Int
    "the vertex index of node B"
    nodeB::Int
    "the state being generated"
    pairstate::SymQObj = StabilizerState("ZZ XX")
    # [...] other parameters
end

@resumable function (prot::EntanglerProt)()
    # [...] entanglement-generation logic
end
\end{lstlisting}

Protocol structures are callable, so invoking a protocol instance begins its execution. To include that execution in the discrete-event simulation, the resulting process is registered with the simulation using the \texttt{@process} macro:

\begin{lstlisting}[language=Julia]
# Entangler for the link between nodes 1 and 2
entangler = EntanglerProt(
              sim, net, 1, 2,
              # other configuration options
            )
@process entangler()
\end{lstlisting}

\begin{figure}[tb]
        \centering
        {\includegraphics[width=\linewidth]{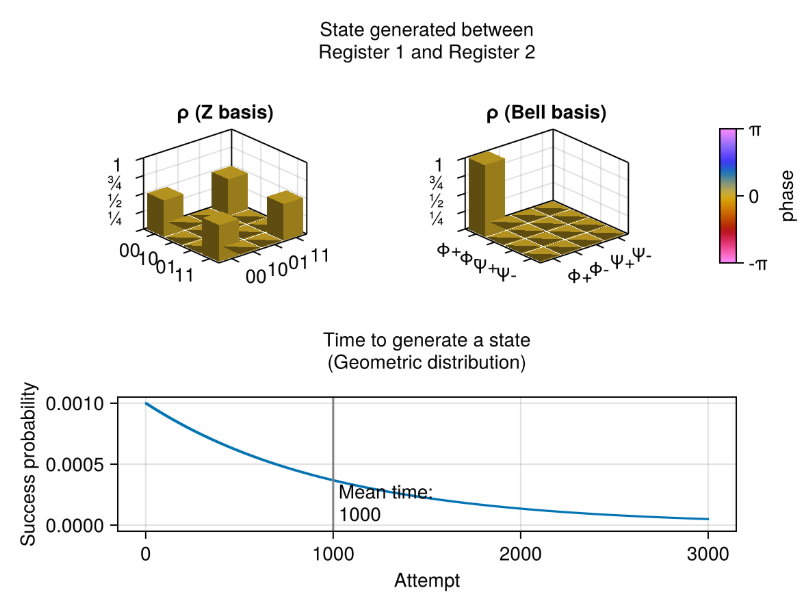}}
        \caption{Protocols in the ProtocolZoo provide specialized visualizations of their internal state and performance, supporting both simulation debugging and analysis of simulation results. The visualization shown here summarizes the entangled state generated by \texttt{EntanglerProt} with its default configuration, together with the probability of successful entanglement generation as a function of the number of attempts.}
        \label{fig:Entanglershow}
\end{figure}

Protocol modules may also define specialized visualization methods for inspecting relevant state and performance metrics. For example, the visualization provided by \texttt{EntanglerProt}, shown in Fig.~\ref{fig:Entanglershow}, displays both the generated quantum state and the probability of successful entanglement generation as a function of the number of attempts.

\subsection{Graphical User Interface}

The unified strictly-defined API of the register object with its universal symbolic representations for states and dynamics and richness of backend simulators, the strict but general tag and query API, the composability of the protocol abstraction, and the StatesZoo, CircuitZoo, and ProtocolZoo repositories make the construction of a graphical user interface (GUI) easy. The simplicity of the GUI codebase is itself a proof of the good composability of the QSavory library -- our APIs follow simple rules and do not introduce bespoke exceptions, thus making them available from a GUI can be done in a fairly automated programmatic fashion instead of requiring manual writing of wrappers. The GUI is publicly available at \href{https://studio.quantumsavory.org}{\url{studio.quantumsavory.org}}, is fully open-source, as is the rest of the library, and has full support for arbitrary network topologies deployed on top of real-world maps, all of the numerical backends, full symbolic support for states and dynamics, and all protocols from the ProtocolZoo. The GUI embeds the library's documentation, providing detailed guidance at each step. A user can use the GUI to understand the basics of the QSavory approach. Moreover, once the user is familiar with the basics and has made a simple project, they can export it as a QSavory script in order to write their own, much more advanced simulation for situations where they need to go beyond the capabilities of the GUI. Lastly, the GUI supports the Model Context Protocol, enabling the user to attach their own AI agents to help them interactively with the construction of more sophisticated simulations.

\section{Full-Stack Examples\label{fullstack}}

This section provides full-stack examples that serve both as technical validation of the abstractions introduced in Sec.~\ref{qsavory} and as implementation templates for building digital twins. Each example couples (i) backend-agnostic, symbolic specifications of states and operations with (ii) a discrete-event execution model for asynchronous LOCC control flow, and (iii) the tag/query metadata plane for resource discovery and coordination across independently developed protocol components. The goal is to show how nontrivial protocol stacks can be expressed without hard-wiring dependencies through explicit handles or bespoke message channels: protocol components synchronize by waiting on, producing, and consuming structured tags attached to registers and message buffers. The examples stress complementary aspects of the framework, including construction and manipulation of multipartite resource states, distributed feed-forward driven by measurement outcomes, long-running concurrent control loops, and contention-aware resource management via locking and querying. Across these layers, physical modeling choices and numerical backends remain swappable without rewriting the protocol logic.

These, and many more such examples are available both in the online documentation and in the code repository of QSavory.

\subsection{Measurement-Based Entanglement Distillation}
\begin{figure}[tb]
        \centering
        Sequence Diagram
        
        \includegraphics[width=0.7\linewidth]{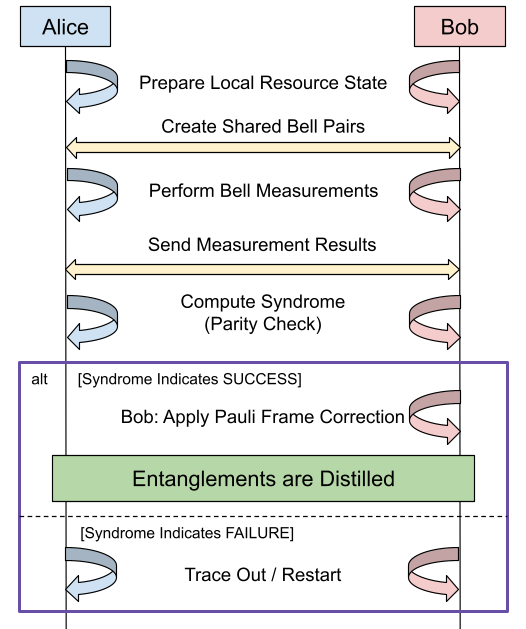}
        \vspace{0.5cm}
        
        Schematic Overview
        \vspace{0.15cm}
        
        \includegraphics[width=\linewidth]{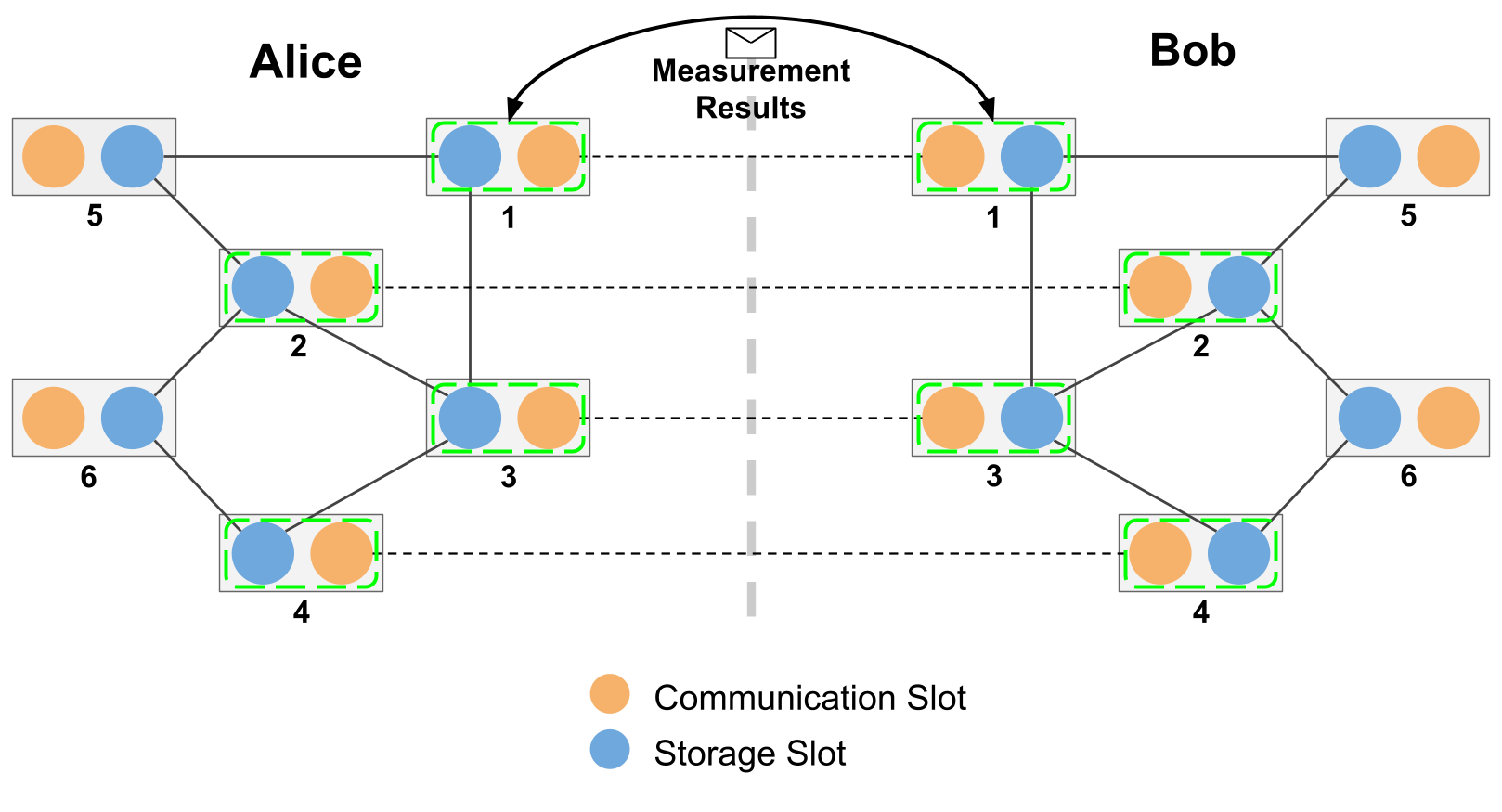}
        \caption{Overview of the protocol using the $[[4,2,2]]$ code. \\ Top: sequence diagram of the classical control flow. \\ Bottom: schematic of the quantum resources, where nodes 1--4 hold the physical qubits of the code and nodes 5--6 hold the auxiliary qubits that become the outputs. Solid lines denote the resource state established in the storage slots (step 1), and black dotted lines represent the noisy Bell pairs to be purified (step 2). Green boxes indicate Bell measurements (step 3), whose outcomes are communicated between Alice and Bob (step 4). If the protocol succeeds, the states held in storage slots of nodes 5 and 6 on the two sides hold the $k=2$ purified Bell pairs.}
        \label{fig:mbqc_422}
\end{figure}

We showcase a full-stack implementation of the MBQC-based entanglement purification (a.k.a.\ distillation) protocol presented in~\cite{shi2025measurement} to highlight QSavory's ability to (i) express MBQC primitives (such as graph-state preparation, Bell measurements, and Pauli-frame updates) in a backend-agnostic symbolic layer while (ii) orchestrating LOCC-style classical control through discrete-event, resumable processes and the tag/query messaging plane. The protocol is parameterized by a $[[n,k,d]]$ CSS code: Alice and Bob each locally prepare a $(n+k)$-qubit resource state; consuming $n$ noisy Bell pairs shared between them then yields, upon success, $k$ purified pairs. Here, we model color-center-like hardware: each network node is a register holding a communication qubit (e.g.\ an electron spin) for entanglement generation and a storage qubit (e.g.\ a nuclear spin) for long-term storage. The implementation consists of four main steps (see Fig.~\ref{fig:mbqc_422}).

\begin{enumerate}
    \item \textbf{Resource State Preparation:} Construct the local resource state required for the MBQC operations.

    \item \textbf{Initial Entanglement Generation:} Establish $n$ initial Bell pairs between Alice and Bob in the communication slots.

    \item \textbf{Bell Measurements:} Perform Bell measurements between each initial entangled pair and its corresponding storage-slot qubit, which is part of the resource state.

    \item \textbf{Syndrome Exchange and Error Detection:} Exchange the combined measurement results via classical communication; Alice and Bob compute the syndrome to verify parity. If the syndrome indicates success, Bob applies a set of Pauli operations to obtain purified entangled pairs. Otherwise, both parties trace out their registers to restart the process.
\end{enumerate}

 Each of these steps is implemented as a resumable process within QSavory. We focus on steps 1 and 4, and only briefly describe steps 2 and 3. The full implementation is available in the \texttt{examples/purificationMBQC} directory.

The resource state needed is $\frac{1}{\sqrt{2^{k}}}
\bigotimes_{j=1}^{k}
\left( \, |\bar{0}_{j}\rangle \, |0_{n+j}\rangle 
\;+\; |\bar{1}_{j}\rangle \, |1_{n+j}\rangle \, \right)$, where $\bar{0}_j$ and $\bar{1}_j$ represent the computational bases of the
$j$-th logical qubit, and $0_{n+j}$ and $1_{n+j}$ represent the bases
of the $(n + j)$-th physical qubit.


Entanglement generation occurs in the communication slots, but the resulting states must be moved into the storage slots for long-term storage. Each node has a single communication qubit, so edges of the graph that share a vertex cannot be entangled simultaneously. The graph is therefore built in rounds: a maximum-cardinality matching -- the largest set of edges sharing no common vertex -- is entangled in parallel, fused into the storage qubits, and the freed communication qubits are reused in the next round, until the whole graph is covered. These rounds are computed by a re-entrant step generator, \texttt{graph\_builder}, which can also be informed of which links actually succeeded in a probabilistic setting and re-plans around failed attempts. The graph constructor is invoked independently by Alice and Bob. Once both graphs are initialized, the local Clifford corrections transform them into the final resource state.

To optimize the process, we need a step generator that uses maximum weight matching to find the largest set of edges in the graph that share no common vertices. After entangling these, fusion operations can be performed, and the process is recursively repeated for the remaining edges until the whole graph is covered. The graph constructor will be invoked independently by Alice and Bob. Once both graphs are initialized, the necessary local operations are applied to transform them into the final resource state.

\begin{lstlisting}[language=Julia]
@resumable function (prot::GraphStateConstructor)()
 # [...] unpack constructor fields

    # graph_builder provides us with an optimized sequence
    # of steps to build the local MBQC resource states
    entangling_steps_generator = graph_builder(graph)

    # these are electron and nuclear qubits
    # which we use to prepare and store the resource state
    slots = ...

    # lock all
    @yield reduce(&, [lock(slot) for slot in slots])

    # prepare all the storage qubits
    for n in nodes
        if !isassigned(net[n][storage_slot])
            initialize!(net[n][storage_slot], X1)
        end
    end

    # run multiple rounds of parallel entangling
    # of independent edges
    while true
        # which edges are we entangling in this round
        current_edges = entangling_steps_generator()
        isnothing(current_edges) && break
        processes = []
        # set up an entangler for each edge
        for (i,j) in current_edges
            # construct EntanglerProt for nodes[i] and nodes[j]
            process = @process entangler()
            push!(processes, process)
        end
        # wait on all entanglers
        @yield reduce(&, processes)
        # perform fusion at each communication qubit
        for (i, j) in current_edges
            regA = net[nodes[i]]
            regB = net[nodes[j]]
            Fusion()(regA, regB,
                     communication_slot, storage_slot)
        end
    end

    # unlock all slots now that the process is finished
    for slot in slots
        unlock(slot)
    end

    # save classical metadata (a tag)
    # that marks the resource state preparation as complete
    uuid = rand(Int)
    for (v, n) in enumerate(nodes)
        tag!(net[n][storage_slot], GraphStateStorage, uuid, v)
    end
end
\end{lstlisting}

Step 2 of establishing long-range entanglements is simple; we use \texttt{EntanglerProt} from the ProtocolZoo. 

In step 3, each party performs a Bell measurement: at each of its $n$ nodes, the local half of a noisy shared pair is measured jointly with the corresponding resource-state qubit. Each Bell measurement thus yields one $XX$ and one $ZZ$ outcome bit. The \texttt{PurifierBellMeasurements} protocol performs these measurements across all $n$ nodes, bit-packs the $n$ $XX$ and $ZZ$ outcomes into two integers. To exchange these outcomes, we define a custom tag. As introduced in Sec.~\ref{subsec:tags}, declaring a plain struct is all it takes to establish a new, well-typed classical interface:

\begin{lstlisting}[language=Julia]
struct PurifierBellMeasurementResults
    node::Int
    measurements_XX::Int
    measurements_ZZ::Int
end
\end{lstlisting}

Step 4 must join classical data (i.e., measurement results above) that become available at different simulation times. The \texttt{MBQCPurificationTracker} therefore queries for the local tag and the remote message,  retrying whenever either is missing; once both are present, it XORs the bit strings and applies the parity-check matrices of the code to obtain the syndrome. The same wait-for-local-and-remote idiom underlies the off-the-shelf \texttt{EntanglementTracker} of the ProtocolZoo.
\begin{lstlisting}[language=Julia]
@resumable function (prot::MBQCPurificationTracker)()
    # [...] unpack fields; nodes is the vector of
    # resource-state node indices,
    # and n is the number of initial Bell pairs
    k = length(nodes) - n

    # we send and receive messages at the chief node
    mb = messagebuffer(net, local_chief_idx)

    while true
        # wait for the local measurement results
        local_tag = @yield query_wait(
            net[local_chief_idx],
            PurifierBellMeasurementResults,
            local_chief_idx, W, W)

        # wait for the remote measurement results
        # (filtered by sender)
        msg = @yield querydelete_wait!(mb,
            PurifierBellMeasurementResults,
            remote_chief_idx, W, W)

        local_XX = local_tag.tag[3]
        local_ZZ = local_tag.tag[4]
        remote_XX = msg.tag[3]
        remote_ZZ = msg.tag[4]

        # [...] XOR the bit strings and compute the
        # syndrome from the parity-check matrices

        if all(iszero, syndrome)
            if correct # true for Bob, false for Alice
               # [...] apply Pauli frame corrections
            end
            # tag the purified qubits (each party's nodes
            # are numbered contiguously from its chief)
            for i in n:n+k-1
                tag!(net[local_chief_idx + i][storage_slot],
                    PurifiedEntanglementCounterpart,
                    remote_chief_idx + i, storage_slot)
            end
        else # distillation failed
            # [...] untag and trace out all involved qubits
        end
    end
end
\end{lstlisting}

If the syndrome check passes, Bob applies the appropriate Pauli operations to his qubits; otherwise, both sides discard their states for a subsequent purification attempt.

This example showcases two things that are hard to do in other simulators (Fig.~\ref{fig:comparison}): building a multipartite graph state rather than just Bell pairs, and composing independent, multi-round protocols through tags and messages. Existing tools typically hard-code the Bell pair as the unit of quantum state and wire protocols together through explicit references to one another, so both tasks would require modifying the simulator core rather than the simulation script.

Furthermore, incorporating more realistic hardware requires no changes to the protocol code above: background noise is declared where the registers are constructed (e.g., \texttt{Register(2, T1Decay(t1))} adds memory decoherence to every slot), and the quality and generation rate of the input pairs are configuration parameters of the entanglers. In particular, the \texttt{pairstate} argument accepts arbitrary symbolic or numeric states, so the idealized Werner-state inputs used above can be replaced by hardware-derived states from the StatesZoo, such as the ZALM source model~\cite{zalm}, while every protocol listing in this section remains untouched.

\subsection{Connectionless Quantum Network}
\begin{figure*}
    \centering
    \includegraphics[width=.95\linewidth]{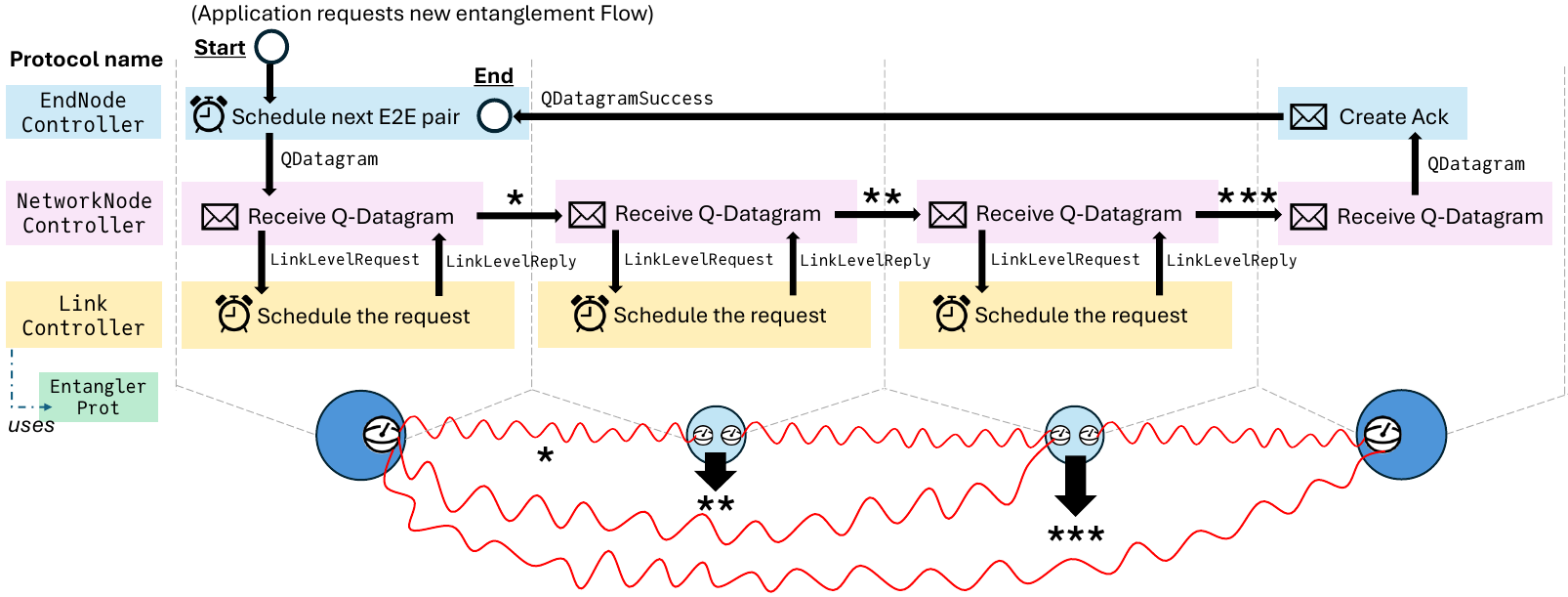}
    \caption{High-level view of the Quantum TCP protocols and their interactions that implement the connectionless architecture. Solid arrows represent classical information being shared between protocols through message buffers, using the specified tags. The bottom part of the figure shows logical snapshots of the entangled state, with wavy red lines representing entanglement, Bloch spheres representing register slots, and blue circles representing registers. We use stars to match one wavy line (i.e., an entangled state between two specific nodes) to a specific step in the interaction diagram above. The logical sequence starts when a new application puts a new \texttt{Flow} tag into the source node's message buffer (\textit{Start} point in the figure). The \texttt{EndNodeController} catches this tag and puts the first \texttt{QDatagram} tag into the same node's message buffer. The \texttt{NetworkNodeController} running at the source node catches the \texttt{QDatagram}, and puts a \texttt{LinkLevelRequest} into, once again, the node's message buffer. This request is caught by the \texttt{LinkController} that takes care of entangling a slot at the source node with a slot of the following register on the chain. When the entanglement is ready, the link controller puts a \texttt{LinkLevelReply} in the buffer. The \texttt{NetworkNodeController} catches it and reacts by forwarding the \texttt{QDatagram} to the following register's message buffer. The handling of the \texttt{QDatagram} is identical at each hop and is repeated until the message reaches the destination node, with the difference that intermediate registers also apply an entanglement swap circuit before forwarding the \texttt{QDatagram}. When the message reaches the destination node's message buffer, it is caught by the \texttt{EndNodeController} instance running on that node, which sends a \texttt{QDatagramSuccess} back to the source node's buffer. This concludes the generation and heralding of an end-to-end entangled pair. Notably, the \texttt{EndNodeController} can generate multiple \texttt{QDatagram}s over time at its discretion, and this interaction loop is repeated independently for each.}
    \label{fig:qtcp}
\end{figure*}

 In many simulations, we need to model full-stack quantum networks in which the quantum data plane is coordinated by classical control protocols responsible for node synchronization, resource allocation, and path selection. 

In this section, we demonstrate how QSavory can model and evaluate a quantum network that serves many concurrent users. We focus on the connectionless two-way architecture introduced in \cite{Bacciottini_2025}, in which internal nodes maintain no per-user state and no resources are reserved ahead of time. Instead, entanglement swapping is performed hop by hop, one link at a time, along the path connecting the communicating users. Figure \ref{fig:qtcp} summarizes how this architecture is implemented in QSavory.

We employ a top-down description and start by illustrating how the entire service can be instantiated through the \texttt{ProtocolZoo} API, which allows users to set up the network as a black-box:

\begin{lstlisting}
net = RegisterNet(...) # the quantum network
endnodes = [...] # a subset of the register indices

for node in endnodes
    end_ctrl = EndNodeController(net, node)
    @process end_ctrl()
end  # all end nodes run an end node controller

for node in vertices(net.graph)
    net_ctrl = NetworkNodeController(net, node)
    @process net_ctrl()
end  # all nodes run a network controller

for edge in edges(net.graph)
    link_ctrl = LinkController(net, edge.src, edge.dst)
    @process link_ctrl()
end  # all links run a link controller.

# a Flow is an intent between any two end nodes:
flow = Flow(src=endnodes[1], dst=endnodes[2],
             npairs=99, uuid=101)
# i.e. endnodes[1] and endnodes[2] want to share 99 Bell pairs

put!(net[endnodes[1]], flow) # EndNodeController handles it

# [...] define as many flows as needed
\end{lstlisting}

The code above applies to any topology and instantiates the full end-to-end entanglement service: once the controllers are running, the network autonomously generates and delivers Bell pairs for all declared flows. On top of this service, users can implement custom applications---such as QKD or other distributed quantum protocols---without modifying the underlying control logic.

We spell out the controllers and their message-level interactions not because users must always work at this level, but because these boundaries are useful extension points. The same simulation can be treated as an opaque end-to-end entanglement service, while still allowing a user to replace one controller, inspect or consume the intermediate messages, or run additional protocols through the shared tag and message-buffer interface. This exposes both vertical extensibility, through specialization of a protocol layer, and horizontal extensibility, through coexistence with other protocol processes in the same network simulation.

The core message of the architecture is the \texttt{QDatagram}, which carries the logical state of a Bell-pair half as it is teleported from the flow source toward its destination. Each \texttt{QDatagram} is associated with a specific flow and encodes metadata such as the flow UUID, sequence number, and accumulated Pauli-frame correction. As swaps are executed, the \texttt{QDatagram} is updated and forwarded hop by hop along the path as in Fig.~\ref{fig:qtcp}. When it finally reaches the destination end node, an end-to-end Bell pair has been successfully established.

\begin{lstlisting}
struct QDatagram
    flow_uuid::Int
    "the flow src, who also creates the qdatagrams"
    flow_src::Int
    "the destination node for the flow"
    flow_dst::Int
    "the Pauli frame correction for the Bell pair"
    correction::Int
    "sequence number of the qdatagram in the given flow"
    seq_num::Int
    "original start time of the qdatagram"
    start_time::Float64
end
\end{lstlisting}

The \texttt{EndNodeController} regulates the number of \texttt{QDatagram}s that each flow injects into the network by maintaining a per-flow congestion window. As illustrated in Fig.~\ref{fig:qtcp} (top left), the controller delays the creation of new end-to-end Bell-pair attempts if the window is full. This prevents intermediate registers from running out of available slots and provides a simple congestion control mechanism compatible with the connectionless architecture. The main logic of this controller is shown below:

\begin{lstlisting}
@resumable function (prot::EndNodeController)()
    (;sim, net, node) = prot
    mb = messagebuffer(net, node)

    # the uuids of flows currently being processed
    current_flows = Set{Int}()

    # [...] Some variables keyed by uuid storing flows data
    # such as the flow destinations, the number of Bell pairs left to deliver, and the windows (see below)

    # The maximum number of qdatagrams in flight per flow
    windows = Dict{Int,Int}()

    while true
        # 1) a new Flow is created ...
        flow = querydelete!(mb, Flow, node, W, W, W)
        #[...] store the flow information in the dictionaries
    
        # 2) received Q-Datagram for which we are the destination ... 
        qdatagram = querydelete!(mb, QDatagram, W, W, node,
                                     W, W, W)
        # ... We send an acknowledgment to the flow source
        ack = QDatagramSuccess(flow_uuid, seq_num, start_time)
        put!(channel(net, node=>flow_src), ack)

        # 3) received an acknowledgment from the dst node... 
        success = querydelete!(mb, QDatagramSuccess, W, W, W)
        # ... retrieve the register slot
        # ... delete flow if we delivered all the Bell pairs
        # ... notify consumers (i.e., send a self message)
        # ... possibly, update window for this flow
        
        # Always: generate as many new QDatagrams as possible
        for uuid in current_flows
            while qdatagrams_in_flight[uuid] < windows[uuid]
                # [...] init fields
                start_time = now(sim)
                qd = QDatagram(
                    uuid, node, dst, corrections, seq_num, start_time,
                )
                put!(net[node], qd)
                qdatagrams_in_flight[uuid] += 1
            end
        end

        # wait for new messages
        @yield onchange(mb)
        
    end
end
\end{lstlisting}

    Because the architecture is connectionless, internal network nodes maintain no per-flow state and execute a minimal, reactive control loop. The \texttt{NetworkNodeController} therefore performs only two operations (Fig.~\ref{fig:qtcp}, pink processing steps): upon receiving a \texttt{QDatagram}, it determines the next hop and issues a \texttt{LinkLevelRequest}; once the corresponding link-level Bell pair is ready (incoming \texttt{LinkLevelReply}), it performs the entanglement swap and forwards the updated \texttt{QDatagram} to the next node.

\begin{lstlisting}
@resumable function (prot::NetworkNodeController)()
    (;sim, net, node) = prot
    mb = messagebuffer(net, node)
    datagrams_in_waiting = ... # keyed by uuid, seq_num; storing datagrams
    while true
        # 1) received a QDatagram ...
        qd = querydelete!(mb, QDatagram, W, W, !=(node),...)
        nexthop = ... # use Graphs.jl to find it
    
        # ...store the QDatagram
        datagrams_in_waiting[(uuid, seq_num)] = qd.tag
    
        # ...request a Bell pair between this node and next hop
        request = LinkLevelRequest(uuid, seq_num, nexthop)
        put!(mb, request)
    
        # 2) new Bell pair between this node and next hop...
        llreply = querydelete!(mb, LinkLevelReply, W, W, W)
        # ...find the QDatagram that matches this reply
        uuid, seq_num, ..., slot_A = llreply.tag
        qd = pop!(datagrams_in_waiting, (uuid, seq_num))
        flow_src = qd[3] # tag head is first; flow_src is third
    
        # [...] some checks (e.g., ensure we are not the flow destination)
    
        # ... entanglement swapping
        if node != flow_src
            # [...] find slot_B associated with the QDatagram
            swapcircuit = LocalEntanglementSwap() # CircuitZoo
            reg = net[node]
            x,z = swapcircuit(reg[slot_A], reg[slot_B])
        end
    
        # ...update and forward QDatagram to next hop
        # [...] compute updated Pauli frame correction
        new_qd = QDatagram(..., new_correction, ...)
        put!(
            channel(net, node => nexthop; permit_forward=false),
            new_qd,
        )
    
        # wait for new messages
        @yield onchange(mb)
    end
    
end
\end{lstlisting}

The \texttt{LinkController} implements the link layer by instantiating \texttt{EntanglerProt} processes on demand. It listens for \texttt{LinkLevelRequest} messages from either endpoint of the link and, upon receiving one, launches an \texttt{EntanglerProt} with the appropriate hardware parameters. When the entanglement attempt succeeds, the controller returns the allocated register slots to the requesting node via a \texttt{LinkLevelReply}, and notifies the opposite endpoint using a \texttt{LinkLevelReplyAtHop}.

\begin{lstlisting}
@resumable function (prot::LinkController)()
    (;sim, net, nodeA, nodeB) = prot
    mbA = messagebuffer(net, nodeA)
    mbB = messagebuffer(net, nodeB)

    while true
        # 1) new request at nodeA...
        llrequest = querydelete!(mbA, LinkLevelRequest, W, W, nodeB)
        _, flow_uuid, seq_num, remote_node = llrequest.tag
        entangler = EntanglerProt(;
            sim, net, nodeA, nodeB, tag=nothing,
            ... # hardware arguments
        )
        proc = @process entangler()
        _, slotA, _, slotB = @yield proc
        # ...send the reply to requesting node
        reply = LinkLevelReply(flow_uuid, seq_num, slotA)
        put!(net[nodeA], reply)

        # send another message type to the other node
        # used by the NetworkNodeController (line 28)
        o_reply = LinkLevelReplyAtHop(flow_uuid, seq_num, slotB)
        put!(net[nodeB], o_reply)

        # 2) new request at nodeB... 
        llrequest = querydelete!(mbB, LinkLevelRequest, W, W, nodeA)
        # [...] same as the other case with flipped replies
        
        # ...wait until we have received a message
        @yield (onchange(mbA) | onchange(mbB))
    end
end
\end{lstlisting}

We note that each of the protocols described above can be replaced independently, provided the substitute exposes the same external interface and message-level API. This modular structure enables users to experiment with alternative congestion-control policies, swapping strategies, or link-level entanglement models without modifying the rest of the stack. The design is therefore fully compatible with the protocol-stack approach presented in \cite{Bacciottini_2025} and commonly adopted in quantum-network architectures.

Finally, we point out that the code snippets shown in this section are meant to provide a conceptual walkthrough of the logic of this architecture. The source code has additional tags and control logic that we did not include for the sake of illustration. We encourage the readers to explore the documentation and the source code for the full implementation.

\subsection{Assisted Quantum Teleportation Network}
To illustrate the flexibility of QuantumSavory in selecting backend representations for quantum states and operations, we consider an assisted continuous-variable quantum teleportation protocol introduced in Ref.~\cite{pirandola2006teleportation}. Restricting to the teleportation of Gaussian states \cite{weedbrook2012gaussian} enables an efficient characterization of qmodes in terms of their first and second statistical moments. In this section, we show how QuantumSavory specifies the full protocol at the level of symbolic qmodes. These formalism-agnostic descriptions can be executed by a numerical backend such as Gabs \cite{kille2025gabs}, enabling fast multi-mode dynamics that avoid the exponential scaling of many-body quantum systems.

Here, we model the quantum teleportation of an input state from Alice to Bob, with the help of a third party, Charlie. While the architecture naturally generalizes to an arbitrary number of assisters, we focus on a single assister for simplicity. The configuration settings can be specified as follows:
\begin{lstlisting}
struct AssistedTeleport <: AbstractProtocol
    sim::Simulation
    net::RegisterNet
    nodeA::Int # Alice (sender)
    nodeB::Int # Bob (receiver)
    nodeC::Int # Charlie (assister)
end
\end{lstlisting}
The first step is to initialize the tripartite entanglement required for teleportation in the \texttt{prepare\_states!} function. This can be accomplished by calling symbolic operations, such as \texttt{SqueezedState} and \texttt{BeamSplitterOp}, that dispatch to Gabs for numerical simulation. It initializes the arbitrary input Gaussian state in Alice's second register slot and generates the shared resource state by mixing local squeezed states via a sequence of beam splitters.
\begin{lstlisting}
function prepare_states!(net, input_state; squeezes)
    # [...]
    # insert "unknown" initial state into Alice's register
    initialize!(regA, 2, input_state)
    # prepare 3-mode EPR state between a, b, c
    initialize!(regA, 1, SqueezedState(-squeezes[1]))
    initialize!(regB, 1, SqueezedState(squeezes[2]))
    initialize!(regC, 1, SqueezedState(squeezes[3]))
    apply!([regA[1], regB[1]], BeamSplitterOp(2/3))
    apply!(regB[1], PhaseShiftOp(1.0 * pi))
    apply!([regB[1], regC[1]], BeamSplitterOp(1/2))
    apply!(regC[1], PhaseShiftOp(1.0 * pi))
    # mix a and input
    apply!(regA[1:2], BeamSplitterOp(1/2))
end
\end{lstlisting}
For brevity, this setup directly prepares a state spanning several nodes; it is a nonlocal simulation shortcut rather than a realizable network protocol. A network-level implementation would establish the resource through explicit entanglement distribution followed by local operations and classical communication.

In the standard continuous-variable teleportation scheme, Alice performs a Bell-like measurement on her local modes. Within the Gaussian formalism, this is implemented as a dual-homodyne detection, that is, Alice measures the position quadrature $q$ of one mode and the momentum quadrature $p$ of the other.
These measurement outcomes encode the classical information required by the receiver to reconstruct the input state. We model this by projecting the register states and transmitting the resulting scalar values through a classical channel to Bob.
\begin{lstlisting}
function homodyne_alice!(net, nodeA, nodeB)
    regA = net[nodeA]
    # project Alice's register onto the eigenstates |x-, p+>
    xminus = project_traceout!(regA, 2, HomodyneMeasurement([0.0]))
    pplus = project_traceout!(regA, 1, HomodyneMeasurement([pi/2]))
    # put quadrature measurements in channel
    chAB = channel(net, nodeA=>nodeB)
    put!(chAB, Tag(:quadsA, -xminus, pplus))
end
\end{lstlisting}
Similarly, the assister, Charlie, performs a homodyne measurement on his register and transmits the outcome to Bob. This operation is quite similar in content to the function \texttt{homodyne\_alice!}, therefore we omit the internal code for \texttt{homodyne\_charlie!} for brevity. With the functions defined, we can assemble the full protocol logic into a resumable function:
\begin{lstlisting}
@resumable function (prot::AssistedTeleport)()
    # [...]
    homodyne_alice!(net, nodeA, nodeB)
    homodyne_charlie!(net, nodeB, nodeC)
    # Bob waits for the classical messages
    # and converts them into his correction.
    regB = net[nodeB]
    mb = messagebuffer(regB)
    quadsAtag = @yield query_wait(mb, :quadsA, W, W)
    quadsA = [quadsAtag.tag[2], quadsAtag.tag[3]]
    quadsCtag = @yield query_wait(mb, :quadsC, W)
    quadC = quadsCtag.tag[2]
    D = DisplaceOp(...) # expression depending on quads
    apply!(regB[1], D)
end
\end{lstlisting}
Here, Bob reconstructs the input state by applying a phase space displacement operator \texttt{DisplaceOp} derived from Alice and Charlie's measurement results. This final operation \texttt{apply!} transforms the state in Bob's register, undoing the random phase space displacement induced by the Bell measurements. Once applied, the input state is recovered in Bob's register.

One of this example's purposes is to show how easy it is to work with arbitrary quantum systems and arbitrary backend simulators thanks to the symbolic frontend and the small universal Register API. Moreover, as a proof of the simplicity afforded by the small universal Register API, we note that only 30 lines of new code needed to be added to QSavory itself and 35 lines of code needed to be added to QuantumSymbolics.jl in order to incorporate Gabs as a backend simulator. This is another proof of the ease with which other simulators can be added in the future.

\section{Related Work: A Cross Comparison\label{sequence}}
We compare QuantumSavory against SeQUeNCe~\cite{wu2021sequence} on a resource-management example that
is deliberately small enough to inspect, yet complex enough to exercise several
independent layers of a quantum-network simulator.  The example combines
memory-resource allocation, Barrett-Kok elementary-link generation \cite{barrett-kok},
entanglement swapping, and BBPSSW \cite{Bennett_1996} distillation.  Agreement on
this example is therefore more informative than agreement on each one of these protocols alone.

The source code and exhaustive documentation on how we implemented, configured, and validated this comparison are available in a supplementary repository~\cite{sequence_qsavory_supplement}.

The scenario is adapted from the SeQUeNCe resource-management
tutorial~\cite{sequence_tutorial}.
It consists of a three-node repeater chain, denoted
$r_1-r_2-r_3$, with midpoint Bell-state measurement stations on the two
elementary links.  Two traffic demands are active at the same time.  The first
flow asks for direct $r_1-r_2$ elementary Bell pairs.  The second flow asks for
end-to-end $r_1-r_3$ pairs, which are created by generating one elementary pair
on each adjacent link and then performing entanglement swapping at $r_2$.
Only these swapped $r_1-r_3$ pairs are then distilled; elementary pairs are never
distilled in this example.  The BBPSSW operation is ideal, with classical
communication delays included but no local gate noise.  Memories are
multiplexed: each reserved pair of
memory slots may attempt elementary-link generation independently and in
parallel with the other reserved slot pairs. We do not model memory
decoherence or memory cutoff, so all observed differences come from generation statistics, protocol timing, and the chosen quantum-state model.

\subsection{Shared Barrett-Kok model}

The Barrett-Kok~\cite{barrett-kok} protocol is an entanglement distribution scheme
that probabilistically generates heralded Bell pairs using two rounds of photon emission/detection.
Both simulators are driven by the same shared Barrett-Kok model:
the elementary success probability, effective attempt time, and raw-state
fidelity are derived once from the common hardware configuration and then
adapted into each simulator.  The exact density matrix used by QuantumSavory's
\texttt{BarrettKokBellPair}, including the definitions of every optical
efficiency and detector parameter, is provided in the supplementary
material~\cite{sequence_qsavory_supplement} for this example.
In addition to this exact model, we instantiate a Werner-state model in
QuantumSavory by using \texttt{DepolarizedBellPair} with the same raw fidelity
as the exact Barrett-Kok pair.  This second model is included because Werner
states are a common approximation in the quantum-networking literature and
because SeQUeNCe is most naturally run in a Bell-diagonal-state representation,
which is Werner-like for this comparison.  Adding the same abstraction to
QuantumSavory gives a direct representation-level validation between the two
simulators, while the exact run separately shows the effect of preserving the
full Barrett-Kok density-matrix structure.

\subsection{Implementation}
\subsubsection{QuantumSavory}
The QuantumSavory implementation is most naturally expressed as a small set of
asynchronous protocols.  We create one register per network node and
connect them with a \texttt{RegisterNet} whose classical delay is derived from
the speed-of-light delay in the shared configuration.  We then start an
\texttt{EntanglementTracker} on every node so that the metadata describing the
remote endpoint of an entangled pair is updated automatically after swaps.

Elementary generation is multiplexed over the reserved memories.  Concretely, the implementation starts an
independent \texttt{EntanglerProt} for each pair of register slots:  
\begin{lstlisting}[language=Julia]
    for (slotA, slotB) in zip(slotsA, slotsB)
        @process EntanglerProt(
            sim, net, nodeA, nodeB;
            chooseslotA=slotA,
            chooseslotB=slotB,
            success_prob=success_prob,
            attempt_time=attempt_time,
            pairstate=pairstate,
        )()
    end
\end{lstlisting}
The \texttt{attempt\_time} and \texttt{success\_prob} values are derived from the shared Barrett-Kok configuration so that they are compatible with the simulated hardware. The keyword arguments \texttt{chooseslotA} and \texttt{chooseslotB} are the
important part of the construction: they bind each entangling process to one
reserved pair of memories, so different reserved memories may attempt
generation concurrently.

The pairstate argument specifies the state produced by each successful entanglement attempt, and is the only part of this configuration that encodes the physical model of the link. Because the downstream protocols consume an abstract two-qubit state, this keeps the physical-layer model independent of the protocol logic. We use two states from the StatesZoo:

\begin{lstlisting}[language=Julia]
BarrettKokBellPair(
    arm_transmissivity,
    arm_transmissivity,
    dark_count_probability,
    efficiency,
    mode_matching_visibility
)

DepolarizedBellPair(;F=barrett_kok_raw_fidelity)
\end{lstlisting}
This is the intended role of \texttt{StatesZoo}: the protocol stack consumes an
abstract two-qubit state, while the physical model of that state can be changed
without rewriting the control flow. Here we have chosen a particular, convenient, level of coarse-graining the simulation, but of course the user might also choose to go deeper and explicitly model lower-level dynamics of creating the Barrett-Kok pair instead of relying on the pre-defined model.

A single \texttt{SwapperProt} runs at the middle
node $r_2$.  It is restricted to the memory ranges reserved for the $r_1$-$r_3$
flow and performs ideal swapping.  Finally, the distillation protocol is installed on
$r_1$ and $r_3$:
\begin{lstlisting}[language=Julia]
@process SwapperProt(
    sim, net, 2;
    nodeL=1, nodeH=3,
    chooseslots=swap_slots,
    local_busy_time=swap_busy,
)()

@process BBPSSWProt(
    sim, net, 1, 3;
    initial_handshake=true, # a preliminary communication round
)()
\end{lstlisting}

\begin{figure*}[t]
    \centering
    \begin{minipage}{0.48\textwidth}
        \centering
        \includegraphics[width=\linewidth]{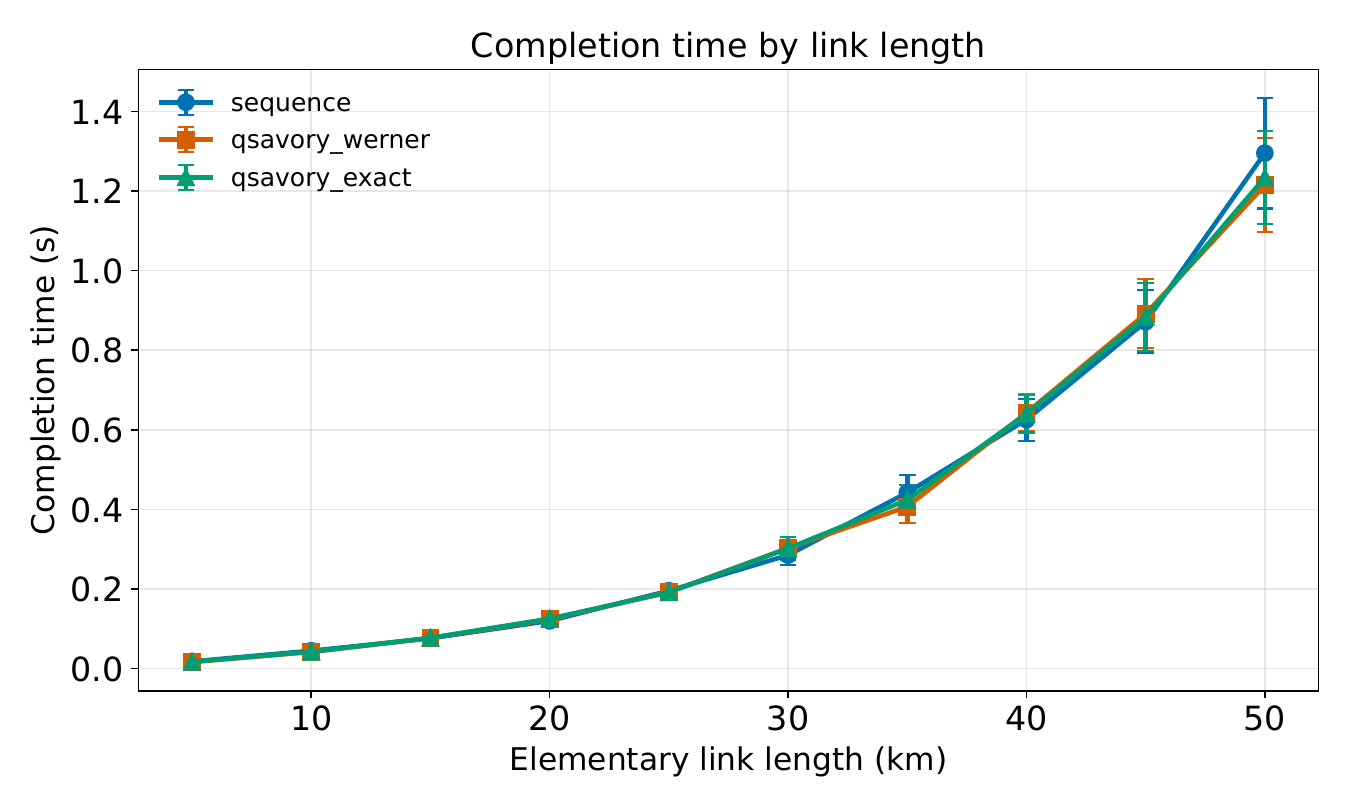}
        \textbf{(a)}\\[0.5ex]
    \end{minipage}
    \hfill
    \begin{minipage}{0.48\textwidth}
        \centering
        \includegraphics[width=\linewidth]{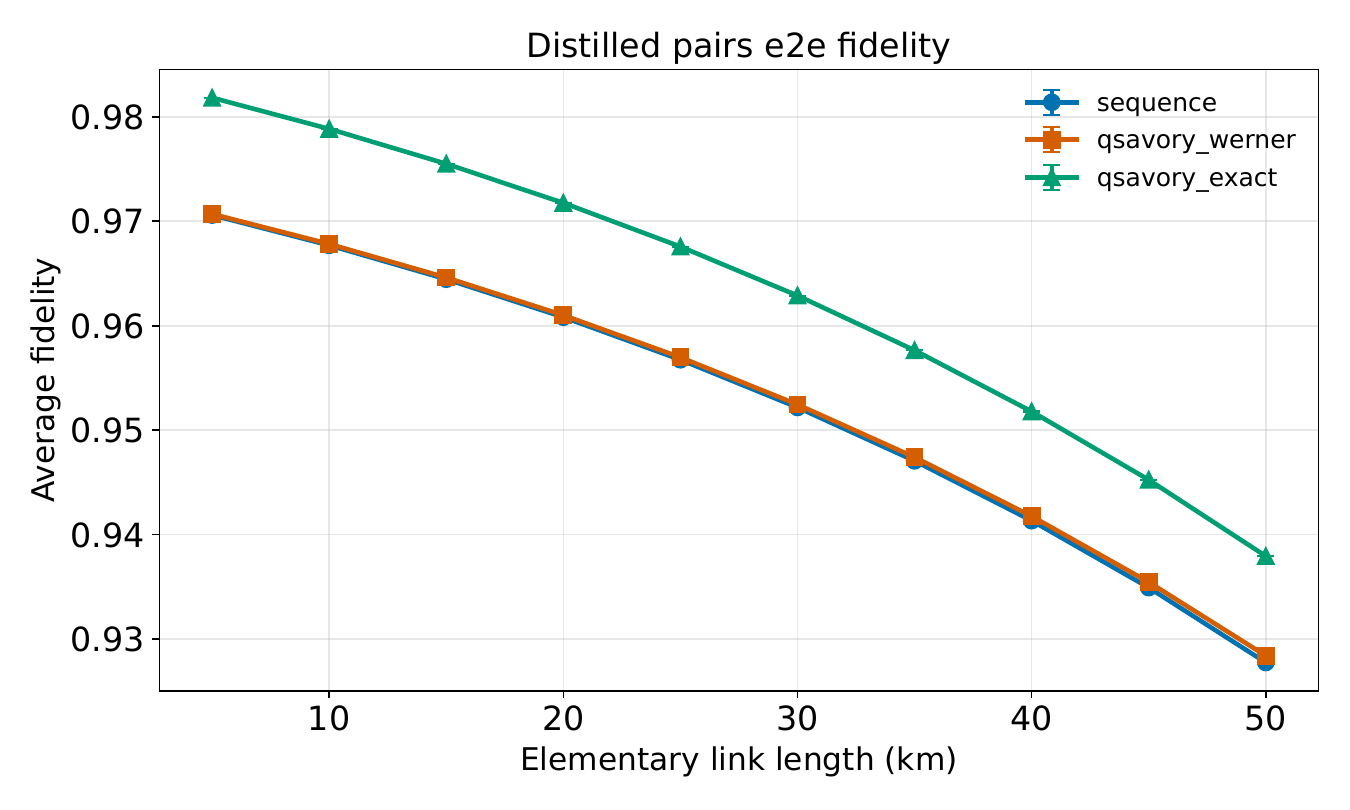}
        \textbf{(b)}\\[0.5ex]
    \end{minipage}
    \caption{Cross-comparison between SeQUeNCe and QuantumSavory over a
    link-length sweep.  (a) Mean completion time for the requested outputs:
    ten $r_1$-$r_2$ pairs and nine distilled $r_1$-$r_3$ pairs.  (b) Mean
    fidelity of distilled end-to-end pairs.  Error bars show $95$\%
    confidence intervals over $30$ independent repetitions.}
    \label{fig:sequence_qsavory_comparison}
\end{figure*}

\subsubsection{SeQUeNCe}

The SeQUeNCe implementation (fully presented in~\cite{sequence_tutorial, sequence_qsavory_supplement}) 
realizes the same shared configuration at a lower
abstraction level.  It instantiates quantum routers, midpoint BSM nodes,
classical and quantum channels, resource-manager rules, Barrett-Kok endpoint
protocols, entanglement-swapping protocols, and BBPSSW distillation rules.  The
same memory reservations determine which memories may participate in each flow,
and purification rules are restricted to swapped $r_1-r_3$ pairs.  SeQUeNCe is
run in its Bell-diagonal-state representation, matching the Werner-like
abstraction used by the \texttt{DepolarizedBellPair} QuantumSavory configuration.

The level of abstraction at which the SeQUeNCe and QSavory simulations are implemented is not the same---we strived to do what is idiomatic in each simulator. Each implementation is documented in detail in the linked reproducible repository~\cite{sequence_qsavory_supplement}.

\subsection{Comparison results}

We sweep the elementary-link length over $5,10,\ldots,50$ km while keeping the
same memory and protocol configuration in both simulators.  The chain has
$20$ register slots at $r_1$, $40$ at $r_2$, and $10$ at $r_3$; ten reserved
slot pairs are available for the direct $r_1$-$r_2$ flow, and ten reserved
slot pairs are available on each elementary link used by the $r_1$-$r_3$ flow.
The elementary-link model uses fiber attenuation $0.2$ dB/km, source
collection efficiency $0.5$, detector efficiency $0.8$, dark-click probability
$5\times 10^{-4}$, and Barrett-Kok visibility $0.99$.  Classical messages are
assumed reliable and delayed only by propagation at $2\times 10^5$ km/s in
fiber.  With these parameters, the Barrett-Kok model gives a
per-reserved-slot elementary generation rate of approximately $473$ Hz at
$5$ km, $191$ Hz at $10$ km, and $6.7$ Hz at $50$ km.  The complete
configuration and the simulator-specific mappings are given in the
supplementary material~\cite{sequence_qsavory_supplement}.

We report two performance measures.  The completion time $T_{\mathrm{comp}}$
is the simulated time at which the experiment has delivered all requested
pairs: ten elementary Bell pairs for the direct $r_1$-$r_2$ flow and nine
distilled Bell pairs for the $r_1$-$r_3$ flow.  The end-to-end fidelity
$F_{\mathrm{e2e}}$ is the
mean fidelity of the distilled $r_1$-$r_3$ pairs delivered by the end-to-end
flow.  Both quantities are averaged over $30$ independent repetitions
for each link length.

Figure~\ref{fig:sequence_qsavory_comparison}a shows that the completion times
of the three simulator series agree within confidence intervals over the full
link-length sweep.

The fidelity comparison in Fig.~\ref{fig:sequence_qsavory_comparison}b
separates two effects.  SeQUeNCe and the QuantumSavory Werner run agree closely
because both use a Bell-diagonal or Werner-like model for the
elementary pair state.  By contrast, the QuantumSavory exact run produces a
visibly different distilled-pair fidelity because \texttt{BarrettKokBellPair}
retains the physical structure of the Barrett-Kok density matrix rather than
compressing it to a single Werner parameter.  In this example, therefore, the
choice of protocol abstraction has no significant effect on the reported
timing metrics, while the choice of quantum-state representation has a much
larger effect on the predicted distilled-pair fidelity.

\section{Conclusion and Future Work\label{conclusions}}
We have presented QSavory, a unified simulation framework designed to support full-stack modeling of quantum computing and quantum networking systems. By combining a symbolic, formalism-agnostic frontend,  
flexible numerical backends, and a discrete-event execution model, QSavory enables users to express complex quantum protocols independently of the underlying simulation representation. Its register abstraction supports heterogeneous quantum systems with declarative noise models, while the tag/query and messaging infrastructure provides a modular mechanism for coordinating classical control, resource management, and protocol composition. Together, these features make it possible to simulate realistic, distributed quantum architectures spanning multiple physical platforms and abstraction layers.

Future work targets scale, accuracy, and reuse of models. A first priority is \emph{surrogate components}: learning a surrogate model from stored runs of an expensive sub-simulation (quantum dynamics plus discrete-event control), in order to create a much more efficient black-box module that reproduces the results of the sub-simulation (success/failure statistics, quantum states, and latency distribution), suitable for embedding in larger simulations; we will support both learned surrogates and reduced-order algorithmic models with explicit uncertainty tracking.

A second priority is adding tensor network backends through the symbolic-frontend-to-backend-simulator interface, including measurement/feed-forward and open-system evolution with controlled truncation and diagnostics. This would greatly expand the type of dynamics that can be modeled with QSavory. Thankfully, many excellent tensor network frameworks already exist to wrap around. Relatedly, we plan to support switching between formalisms \emph{on the fly}.

Third, we will expand and systematize the reusable libraries of states, circuits, and protocol modules (QSavory's \textit{Zoos}) by standardizing interfaces (resource requirements and tag/message schemas) and attaching machine-readable performance metadata to enable drop-in substitution, benchmarking, and visualization. On the physical side, we will develop higher-fidelity in-transit and photonic channel models with explicit mode structure, detection models, multiplexing constraints, and event-driven timing. Finally, we will strengthen support for network error-correction layers (scheduled syndrome workflows, decoder hooks with latency models) and curate databases of entanglement purification circuits.

QSavory also benefits from the development of an open-source graphical user interface that we will continue investing in.
\begin{acknowledgments}
S.K. conceived of the software project and performed much of the early development. H.K., L.B., and A.B. contributed much of the recent development work. L.B. conceived of the qTCP protocol. A.K. contributed the Gaussian-state backend and significant work on the symbolic capabilities. The manuscript was written mainly by H.K. and L.B., with input from all authors. We acknowledge support from NSF Grants No. 1941583, 2346089, 2402861, and 2522101. A.K. acknowledges support from the Pierre Hohenberg Graduate Fellowship provided by the Flatiron Institute. The Flatiron Institute is a division of the Simons Foundation. We are grateful for the helpful input from Don Towsley, Dirk Englund, and Saikat Guha.
\end{acknowledgments}








\vspace{2in}

\bibliography{bibliography}

\end{document}